\documentclass[12pt]{article}
\usepackage[margin=1.25in]{geometry}
\usepackage{natbib}
\usepackage{amsmath}
\usepackage{amssymb}
\usepackage{amsfonts}
\usepackage{multirow}
\usepackage{mathtools}
\usepackage{verbatim}
\usepackage{amsthm}
\usepackage{color}
\usepackage{bm}
\usepackage{float}

\DeclarePairedDelimiter\abs{\lvert}{\rvert}%

\theoremstyle{definition}
\newtheorem{definition}{Definition}

\begin{document}


\renewcommand{\baselinestretch}{2}

\markright{ \hbox{\footnotesize\rm Statistica Sinica
}\hfill\\[-13pt]
\hbox{\footnotesize\rm
}\hfill }

\markboth{\hfill{\footnotesize\rm FIRSTNAME1 LASTNAME1 AND FIRSTNAME2 LASTNAME2} \hfill}
{\hfill {\footnotesize\rm FILL IN A SHORT RUNNING TITLE} \hfill}

\renewcommand{\thefootnote}{}
$\ $\par


\fontsize{12}{14pt plus.8pt minus .6pt}\selectfont \vspace{0.8pc}
\centerline{\large\bf Mechanisms for Global Differential Privacy  }
\vspace{2pt} 
\centerline{\large\bf under Bayesian Data Synthesis}
\vspace{.4cm} 
\centerline{Jingchen Hu, Matthew R. Williams and Terrance D. Savitsky} 
\vspace{.4cm} 
\centerline{\it Vassar College, RTI International and U.S. Bureau of Labor Statistics}
 \vspace{.55cm} \fontsize{9}{11.5pt plus.8pt minus.6pt}\selectfont


\begin{quotation}
\noindent {\it Abstract:}
This paper introduces a new method that embeds any Bayesian model used to generate synthetic data and converts it into a differentially private (DP) mechanism.  We propose an alteration of the model synthesizer to utilize a censored likelihood that induces upper and lower bounds of [$\exp(-\epsilon / 2), \exp(\epsilon / 2)$], where $\epsilon$ denotes the level of the DP guarantee.  This censoring mechanism equipped with an $\epsilon-$DP guarantee will induce distortion into the joint parameter posterior distribution by flattening or shifting the distribution towards a weakly informative prior.   To minimize the distortion in the posterior distribution induced by likelihood censoring, we embed a vector-weighted pseudo posterior mechanism within the censoring mechanism.  The pseudo posterior is formulated by selectively downweighting each likelihood contribution proportionally to its disclosure risk.  On its own, the pseudo posterior mechanism produces a weaker asymptotic differential privacy (aDP) guarantee. After embedding in the censoring mechanism, the DP guarantee becomes strict such that it does not rely on asymptotics.  We demonstrate that the pseudo posterior mechanism creates synthetic data with the highest utility at the price of a weaker, aDP guarantee, while embedding the pseudo posterior mechanism in the proposed censoring mechanism produces synthetic data with a stronger, non-asymptotic DP guarantee at the cost of slightly reduced utility. The perturbed histogram mechanism is included for comparison.

\vspace{9pt}
\par
\end{quotation}\par

\def\thefigure{\arabic{figure}}
\def\thetable{\arabic{table}}

\renewcommand{\theequation}{\thesection.\arabic{equation}}

\fontsize{12}{14pt plus.8pt minus .6pt}\selectfont

\section{Introduction}
\label{sec:intro}

This paper focuses on constructing a synthetic data mechanism equipped with a formal privacy guarantee that allows the use of any Bayesian probability model as the data synthesizer.  A formal privacy guarantee is quantifiable and attaches to a data generating mechanism independent of the behaviors of putative intruders seeking to re-identify data records; in particular, we focus on the differential privacy (DP) framework \citep{Dwork:2006:CNS:2180286.2180305} to provide a formal privacy guarantee.  


\begin{definition}[Differential Privacy]\label{def:DP}
Let $D \in \mathbb{R}^{n \times k}$ be a database in input space $\mathcal{D}$. Let $\mathcal{M}$ be a randomized mechanism such
that $\mathcal{M}: \mathbb{R}^{n \times k} \rightarrow O$. Then $\mathcal{M}$ is $\epsilon$-differentially private if
\[
\frac{Pr[\mathcal{M}(D) \in O]}{Pr[\mathcal{M}(D^{'}) \in O]} \le \exp(\epsilon),
\]
for all possible outputs $O = Range(\mathcal{M})$ under all possible pairs of datasets $D, D^{'} \in \mathcal{D}$ of the same size which differ by only a single row (i.e., Hamming-1 distance).
\end{definition}

DP assigns a disclosure risk for a statistic to be released to the public, $f(D)$ (e.g., total employment for a state-industry) of any $D \in \mathcal{D}$ based on the global sensitivity, $\Delta_G = \mathop{\sup}_{D,D^{'}\in\mathcal{D}: ~\delta(D,D^{'})=1}\abs{f(D) - f(D^{'})}$, over the space of databases, $\mathcal{D}$, where $\delta(D,D^{'}) = 1$ denotes the Hamming-1 distance such that $D$ differs from $D^{'}$ by a single record. If the value of the statistic, $f$, expresses a high magnitude change after changes or the removal of a data record in $D^{'}$ (i.e., large $\Delta_G$ value), then the mechanism will be required to induce a relatively higher level of distortion to $f$.  

A commonly-used data privacy approach generates synthetic microdata from statistical models estimated on confidential data for proposed release by data disseminators \citep{Rubin1993synthetic, Little1993synthetic}. We denote such a release mechanism used to generate synthetic data, $\xi(\theta \mid \mathbf{x})$, where $\mathbf{x}$ are the confidential data and $\theta$ denotes the parameters used to generate the synthetic data. In particular, we consider the case where $\xi(\theta \mid \mathbf{x})$ is a posterior distribution of a Bayesian hierarchical probability model.

The Exponential Mechanism by \citet{McSherryTalwar2007} is a popular approach to generating draws of parameters, $\theta$, and associated synthetic data which is differentially private. 

\begin{definition}[Exponential Mechanism]
The Exponential Mechanism releases values of $\theta$ from a distribution proportional to
\begin{equation}
\exp \left(\frac{\epsilon u(\mathbf{x}, \theta)}{2 \Delta_u} \right)\xi(\theta),
\end{equation}
where $u(\mathbf{x}, \theta)$ is a utility function and $\xi(\theta)$ is a base or prior distribution. Let
\begin{equation}
\Delta_{u} = \mathop{\sup}_{\mathbf{x}\in \mathcal{X}^{n}} \,\, \mathop{\sup}_{\mathbf{x}, \mathbf{y}: \delta(\mathbf{x}, \mathbf{y}) = 1} \, \, \mathop{\sup}_{\theta \in \Theta} \,\, \lvert u(\mathbf{x}, \theta) - u(\mathbf{y}, \theta) \rvert   
\end{equation} 
denote the sensitivity of the utility function, $u(\mathbf{x},\theta)$, defined globally over $\mathbf{x} = (x_{1},\ldots,x_{n}) \in \mathcal{X}^n$, the $\sigma-$algebra of datasets, $\mathbf{x}$, governed by product measure, $P_{\theta_{0}}$; $\delta(\mathbf{x}, \mathbf{y}) = \# \{i: x_i \neq y_i\}$ is the Hamming distance between $\mathbf{x}, \mathbf{y} \in \mathcal{X}^n$. Then each draw of $\theta$ from the Exponential Mechanism provides an $\epsilon-$DP privacy guarantee, where $\epsilon \leq 2\Delta_{u}$.
\end{definition}

For a Bayesian model utilizing the data log-likelihood as the utility function of the exponential mechanism, \citet{SavitskyWilliamsHu2020ppm} demonstrate the exponential mechanism specializes to the model posterior distribution, which provides a straightforward mechanism from which to draw samples. \citet{Dimitrakakis:2017:DPB:3122009.3122020} define a model-based  sensitivity, $ \mathop{\sup}_{\mathbf{x},\mathbf{y}\in \mathcal{X}^{n}:\delta(\mathbf{x}, \mathbf{y}) = 1}  \mathop{\sup}_{\theta \in \Theta} \lvert f_{\theta}(\mathbf{x}) - f_{\theta}(\mathbf{y}) \rvert \le \Delta$, which is constructed as a Lipschitz bound.  They demonstrate a connection between the Lipschitz bound, $\Delta$ and $\epsilon \leq 2\Delta$ for each draw of parameters, $\theta$, where $f_{\theta}(\mathbf{x})$ is the model log-likelihood.  The guarantee applies to all databases $\mathbf{x}$, in the space of databases of size $n$, $\mathcal{X}^{n}$, and denoted as the posterior mechanism. 

Computing a finite $\Delta < \infty$ in practice, however, as acknowledged by \citet{Dimitrakakis:2017:DPB:3122009.3122020}, is difficult-to-impossible for an unbounded parameter space (e.g., a normal distribution) under simple models. A truncation of the parameter space is required to achieve a finite $\Delta$ and the truncation only works for some models to achieve a finite $\Delta$.  Even more, parameter truncation becomes intractable analytically or computationally or both for practical models that utilize a multidimensional parameter space. For example, a non-linear mixed effects model could have the number of model parameters on the order of the number of data records, which can be millions, in practice.

The pseudo posterior mechanism of \citet{SavitskyWilliamsHu2020ppm} reviewed in Section \ref{sec:methods:weighted} allows achievement of a target Lipschitz value that, in turn, determines a formal privacy guarantee by using likelihood weights, $\alpha_{i} \in [0,1]$, to selectively downweight high-risk records.  This mechanism achieves a DP guarantee \emph{asymptotically} as the parameter space contracts onto a point. In other words, while the mechanism is flexible, the privacy guarantee is not strict.

This paper introduces a new mechanism that embeds the pseudo posterior mechanism by \emph{censoring} the pseudo likelihood values above or below a threshold to targeted Lipschitz (and hence privacy) guarantee, which allows achievement of a target $\epsilon-$DP guarantee without reliance on asymptotics.  Any Bayesian data synthesizing model may be utilized under the censoring mechanism.  

The remainder of the paper is organized as the following. Section \ref{sec:methods} reviews the pseudo posterior mechanism, followed by a review of the perturbed histogram mechanism that will be used for comparison. The new censoring mechanism is introduced in Section \ref{sec:censoring}. Section \ref{sec:sim} focuses on a series of simulation studies where the mechanisms are applied and their data utilities are compared at equivalent privacy guarantees. 
Section \ref{sec:app} presents an application to a sample of the Survey of Doctoral Recipients. 
We end the paper with a few concluding remarks in Section \ref{sec:conclusion}.

\section{Synthetic Microdata Generating Privacy Mechanisms}
\label{sec:methods}

In this section, we review the pseudo posterior mechanism proposed by \citet{SavitskyWilliamsHu2020ppm} in Section \ref{sec:methods:weighted} and discuss a variant of it in Section \ref{sec:methods:Weighted-e}. The perturbed histogram mechanism \citep{Dwork:2006:CNS:2180286.2180305, WassermanZhou2010}, 
is briefly reviewed in Section \ref{sec:methods:PH}.

\subsection{Pseudo Posterior Mechanism}
\label{sec:methods:weighted}


To guarantee the achievement of a finite $\Delta < \infty$ for any synthesizing model over an unbounded parameter space, \citet{SavitskyWilliamsHu2020ppm} propose a \emph{pseudo} posterior mechanism that uses a log-pseudo likelihood with a vector of observation-indexed weights $\bm{\alpha} = (\alpha_1, \cdots, \alpha_n) \in [0,1]^{n}$ where each $\alpha_{i}$ exponentiates the likelihood contribution, $p(x_{i}\mid \theta)$, for each record $i \in (1,\ldots,n)$.  Each weight, $\alpha_{i} \in [0, 1]$ is set to be inversely proportional to a measure of disclosure risk for record, $i$, such that the model used to generate synthetic data will be less influenced by relatively high-risk records.  

The pseudo posterior mechanism of \citet{SavitskyWilliamsHu2020ppm} is formulated as
\begin{equation}
\label{pseudomech}
\xi^{\bm{\alpha}(\mathbf{x})}(\theta \mid \mathbf{x}) \propto \mathop{\prod}_{i=1}^{n}p(x_{i}\mid \theta)^{\alpha_{i}} \times \xi(\theta),
\end{equation}
where each $\alpha_{i} \in [0,1]$ serves to downweight the likelihood contribution of record $i$ with $\alpha_{i} \propto 1/\mathop{\sup}_{\theta \in \Theta} \abs{f_{\theta}(x_{i})}$, and $\abs{f_{\theta}(x_{i})} = \abs{\log p(x_{i}\mid \theta)}$ is the absolute log-likelihood of record $i$. The differential downweighting of individual records intends to better preserve utility by focusing the downweighting on high-risk records.  High-risk records tend to be those located in the tails of the distribution where the absolute log-likelihood, $\abs{f_{\theta}(x_{i})}$, is highest. The differential downweighting allows the preservation of the high mass portions of the data distribution in the generated synthetic data.  The pseudo posterior mechanism sets $\alpha_{i} = 0$ for any record with a \emph{non-finite} log-likelihood, which ensures a finite $\Delta_{\bm{\alpha}} = \mathop{\sup}_{\mathbf{x},\mathbf{y}\in \mathcal{X}^{n}:\delta(\mathbf{x}, \mathbf{y}) = 1}  \mathop{\sup}_{\theta \in \Theta} \lvert \alpha(\mathbf{x}) \times f_{\theta}(\mathbf{x}) - \alpha(\mathbf{y}) \times f_{\theta}(\mathbf{y}) \rvert < \infty$,  where $\alpha(\mathbf{x}) = \left[\alpha_{1}(\mathbf{x}),\ldots,\alpha_{n}(\mathbf{x})\right]$, $f_{\theta}(\mathbf{x}) = \left[f_{\theta}(x_{1}),\ldots,f_{\theta}(x_{n})\right]$ and $\alpha(\mathbf{x}) \times f_{\theta}(\mathbf{x}) = \mathop{\sum}_{i=1}^{n} \alpha_{i}(\mathbf{x}) f_{\theta}(x_{i})$ that downweights the associated log-likelihood contributions, $f_{\theta}(x_{i})$ with multiplication by $\alpha_{i}(\mathbf{x}) \leq 1$.  We see that $\Delta_{\bm{\alpha}} \leq \Delta$ since $\alpha_{i}(\mathbf{x}) = \alpha_{i} \leq 1$.  The  $\bm{\alpha}-$weighted pseudo synthesizer, $\xi^{\bm{\alpha}(\mathbf{x})}(\theta \mid \mathbf{x})$ satisfies $\epsilon-$DP if the following inequality holds.
\begin{definition}[Differential Privacy under the Pseudo Posterior Mechanism]\label{def:DP-ppm}
\begin{equation}
\mathop{\sup}_{\mathbf{x},\mathbf{y}\in \mathcal{X}^{n}:\delta(\mathbf{x}, \mathbf{y}) = 1} \mathop{\sup}_{B \in \beta_{\Theta}} \frac{\xi^{\bm{\alpha}(\mathbf{x})}(B \mid \mathbf{x})}{\xi^{\bm{\alpha}(\mathbf{y})}(B \mid \mathbf{y})} \leq e^{\epsilon}, 
\end{equation}
where $\xi^{\bm{\alpha}(\mathbf{x})}(B \mid \mathbf{x}) = \int_{\theta\in B}\xi^{\bm{\alpha}(\mathbf{x})}(\theta \mid \mathbf{x})d\theta$.
\end{definition}
 Definition~\ref{def:DP-ppm} limits the change in the pseudo posterior distribution over all sets, $B \in \beta_{\Theta}$ (i.e., $\beta_{\Theta}$ is the $\sigma-$algebra of measurable sets on $\Theta$), from the inclusion of a single record.   Although the pseudo posterior distribution mass assigned to $B$ depends on $\mathbf{x}$, the $\epsilon$ guarantee is defined as the supremum over all $\mathbf{x} \in \mathcal{X}^{n}$ and for all $ \mathbf{y}\in \mathcal{X}^{n}$ which differ by one record (i.e., $\delta(\mathbf{x},\mathbf{y})=1$).
 
Let $\Delta_{\bm{\alpha},\mathbf{x}} = \mathop{\sup}_{\delta(\mathbf{x}, \mathbf{y}) = 1}  \mathop{\sup}_{\theta \in \Theta} \lvert \alpha(\mathbf{x}) \times f_{\theta}(\mathbf{x}) - \alpha(\mathbf{y}) \times f_{\theta}(\mathbf{y}) \rvert $ be the Lipschitz bound computed, \emph{locally}, on database $\mathbf{x}$, over all databases, $\mathbf{y}$, at a Hamming-1 distance from $\mathbf{x}$. (We note that the term local in the DP literature can refer to the trust model as in central versus local or the sensitivity calculation as in global versus local. Our use here, however, differs from these two and focus on being local to the database $\mathbf{x}$.) The pseudo posterior mechanism \emph{indirectly} sets the local DP guarantee, $\epsilon_{\mathbf{x}} = 2 \Delta_{\bm{\alpha},\mathbf{x}}$, through the computation of the likelihood weights, $\bm{\alpha}$. The $\bm{\alpha}$ may be further scaled and shifted; for example, tuning settings $(c_1, c_2)$ can be used to construct $\tilde{\alpha}_{i} = c_1 \times \alpha_{i} + c_2$ and decrease or increase the $\tilde{\bm{\alpha}}$ weights to achieve a target $\Delta_{\tilde{\bm{\alpha}},\mathbf{x}}$, which in turn delivers a target $\epsilon_{\mathbf{x}} = 2 \Delta_{\tilde{\bm{\alpha}},\mathbf{x}}$ \citep{SavitskyWilliamsHu2020ppm, HuSavitskyWilliams2021tabular}.

\citet{SavitskyWilliamsHu2020ppm} show that the local $\Delta_{\bm{\alpha},\mathbf{x}}$ estimated on a confidential database, $\mathbf{x}$, contracts onto the global $\Delta_{\bm{\alpha}}$ over the space of databases of size $n$ asymptotically in sample size, $n$; in particular, $\Delta_{\bm{\alpha},\mathbf{x}}$ becomes arbitrarily close to the global $\Delta_{\bm{\alpha}}$ such that the pseudo posterior mechanism is asymptotically differentially private (aDP).  

The process of creating synthetic data under the $\bm{\alpha}-$weighted pseudo posterior mechanism works as follows:
\begin{enumerate}
\item For a confidential database $\mathbf{x}$ of $n$ records, one fits a Bayesian model parameterized by $\theta$. 
\item Based on the estimated model and the log-likelihood of each record, one calculates a vector of observation-indexed weights $\bm{\alpha} = (\alpha_1, \cdots, \alpha_n) \in [0, 1]^n$ where $\alpha_{i} \propto 1/\abs{f_{\theta}(x_{i})}$.
\item Re-estimate the model under the pseudo posterior in Equation~(\ref{pseudomech}). The local Lipschitz bound, $\Delta_{\bm{\alpha}, \mathbf{x}}$, is computed based on the estimated $\theta^{\ast}$ under the pseudo posterior. These data satisfy a \emph{local} $(\epsilon_{\mathbf{x}} = 2 \times \Delta_{\bm{\alpha}, \mathbf{x}})-$DP guarantee that contracts on an $\epsilon = \epsilon_{\mathbf{x}}$ global aDP guarantee for $n$ sufficiently large.
\item As a post-processing step, a synthetic dataset $\mathbf{x}^*$ is generated from the pseudo posterior predictive distribution under estimated parameters $\theta^*$.  
\end{enumerate}

We label the $\bm{\alpha}-$weighted pseudo posterior mechanism as ``Weighted" in the simulations and application in Sections \ref{sec:sim} and \ref{sec:app}, as contrasted with an ``Unweighted" method that sets all of the $\alpha_{i} = 1$ to produce a posterior mechanism.

\subsection{Weighted-e Pseudo Posterior Mechanism}
\label{sec:methods:Weighted-e}

In addition to the pseudo posterior mechanism with observation-indexed weight $\alpha_i$, \citet{SavitskyWilliamsHu2020ppm} introduce a truncation of each weight, $\alpha_{i}$: if its log-likelihood contribution, $\alpha_{i} \times f_{\theta}(x_i) > \epsilon/2$, we set final weight, $\alpha^{\ast}_{i} = 0$.  

Setting a likelihood threshold, $\epsilon/2$, reduces the constant of proportionality in the $\mathcal{O}(n^{-1/2})$ contraction of the local Lipschitz bounds to speed convergence onto $\Delta_{\bm{\alpha}} = \epsilon/2$.  This adjustment to the pseudo posterior mechanism induces a rapid contraction of the $\epsilon_{\mathbf{x}}$ computed on the observed (local) database to the global $\epsilon$ providing the aDP guarantee.   We include this tweak to the pseudo posterior mechanism because it provides an alternative to the censoring mechanism introduced in the next section for achieving a stronger privacy guarantee.

The synthetic data generation process works in the same way with the $\bm{\alpha}-$weighted pseudo posterior mechanism in Section \ref{sec:methods:weighted}, excepts for their slight differences in how the weights are calculated. We label this slightly revised version of the pseudo posterior mechanism as ``Weighted-e" in the sequel.

\subsection{Perturbed Histogram Mechanism}
\label{sec:methods:PH}
We next present the commonly-used perturbed histogram mechanism for simulating synthetic microdata that achieves $\epsilon-$DP guarantee \citep{Dwork:2006:CNS:2180286.2180305, WassermanZhou2010} as a comparison.  For brevity, we include a detailed review in the Supplementary Materials. 

Under the required strong assumption of a \emph{bounded} and continuous variable, one first discretizes it into a histogram with a selected number of bins. One next induces a formal $\epsilon-$DP privacy guarantee into the histogram by adding Laplace noise. Finally, one simulates microdata from the private histogram under $\epsilon-$DP, which is a post-processing step in a similar fashion as generating synthetic data under the pseudo posterior mechanism (given the privacy protected parameter draws) reviewed in Section \ref{sec:methods:weighted}. We label the perturbed histogram synthesizer as ``PH" in the sequel.

\section{Censored (Likelihood) Mechanism}
\label{sec:censoring}
The pseudo posterior mechanism discussed in Section~\ref{sec:methods:weighted} solves the important problem of how to turn \emph{any} Bayesian probability model into a data synthesizing mechanism equipped with a formal privacy guarantee.   A substantial limitation, however, is that the formal privacy guarantee is asymptotic, not strict.  While an asymptotic guarantee may be favorably viewed as ``relaxed" in the sense that better utility may be achieved than under a strict guarantee, one may never know with certainty when their database size, $n$, is sufficiently large such that the asymptotic privacy guarantee de facto becomes strict (or global over the space of databases). 

Our new mechanism censors the log-likelihood contribution in a manner that carries over the property of the pseudo posterior mechanism of \citet{SavitskyWilliamsHu2020ppm} that allows any Bayesian probability model to be purposed as a formally private mechanism for generating synthetic data, but is now equipped with a strict $\epsilon-$DP guarantee.  It is common practice to construct a Bayesian hierarchical probability model to closely fit the data for the purpose of removing measurement error as is done in small area estimation models estimated on area-indexed survey statistics.   Our newly proposed censoring mechanism makes it straightforward to embed this type of closely-fitting model and yet output synthetic data with a strict $\epsilon-$DP guarantee.

The idea of censoring contributions to log-likelihoods has also appeared in \cite{PenaBarrientos2021} and \cite{Cannone2019ACM} in the DP literature.  Both papers are focused on the narrower task of conducting differentially private simple hypothesis testing.  \cite{PenaBarrientos2021} censors a Bayes Factor used to conduct differentially private hypothesis testing between two model alternatives.   While they use fully Bayesian model specifications, simple conjugate frameworks are used such that the parameters may be analytically marginalized to compute the Bayes factor.   Their example is a nested linear regression under a mixture of g-priors for the regression coefficient.  \cite{Cannone2019ACM} performs differentially private hypothesis testing to differentiate distributions $P$ and $Q$ by censoring or “clamping” the log-likelihood ratio. 

Our new censoring mechanism, by contrast, outputs differentially private model parameters that facilitate synthetic data generation.   Such data may, in turn, be used for many purposes, including hypothesis testing between any number of model parameterization alternatives.   Even more, our mechanism may be used to privatize complex hierarchical models where parameter posterior distributions are not analytically marginalized. We note that the DP methods from \cite{PenaBarrientos2021} and \cite{Cannone2019ACM} are tailored for the specific task of conducting differentially private simple hypothesis testing. Compared to our goal of generating differentially private synthetic data with more flexibility in reusing the outputs for multiple tasks, these specific task oriented DP mechanisms may have the advantage of offering improved utility of the targeted analyses at the expense of less flexibility.  

We demonstrate in the sequel that our new censoring mechanism may even embed the pseudo posterior mechanism with the result that one gets the best properties of both mechanisms.  The distribution properties of the confidential data are well-preserved in the resulting synthetic data due to the surgical targeting of only high-risk record for likelihood downweighting.  Yet, now the resulting synthetic data are equipped with a strict $\epsilon-$DP privacy guarantee.

We begin by laying out the details of how to construct the new censoring mechanism achieving a strict $\epsilon-$DP guarantee as an alternative to relying on the asymptotic contraction of $\theta\in\Theta$ (at $\theta^{\ast}$) to achieve an aDP global privacy guarantee for a sufficiently large sample size, $n$.  Our new mechanism censors the log-likelihood at a target threshold, $\epsilon/2$, representing the targeted DP guarantee.  The use of censoring intends to ``lock in" any chosen $\epsilon$ under an $\epsilon-$DP formal privacy guarantee without relying on asymptotics.  The $\epsilon$ guarantee is strict.  The censored threshold will determine the Lipschitz bound, $\Delta = \epsilon/2$, which will indirectly set the $\epsilon = 2\Delta$ guarantee.  

The definition of DP anchors its guarantee to all databases of fixed size $n$.  By contrast, the privacy guarantee offered by the new censoring mechanism is stronger in that the same $\epsilon$ guarantee applies to databases of any size, including those $>n$.

We introduce the censoring mechanism by embedding the $\bm{\alpha}-$weighted pseudo likelihood to construct the following new pseudo likelihood with,
\begin{equation}
p^{\bm{\alpha}}_c(x_i \mid \theta) = \left\{
\begin{array}{ll}
\exp(\epsilon/2), & p(x_i \mid \theta)^{\alpha_i} > \exp(\epsilon/2),\\
\exp(-\epsilon/2), & p(x_i \mid \theta)^{\alpha_i} < \exp(-\epsilon/2),\\
p(x_i \mid \theta)^{\bm{\alpha}}, & \text{otherwise},
\end{array}
\right.
\label{eq:censor_like}
\end{equation} 
for use in
\begin{equation}
\xi^{\bm{\alpha}}_{c} (\theta \mid \mathbf{x}) \propto \prod_{i=1}^{n} p^{\bm{\alpha}}_{c}(x_i \mid \theta)  \xi(\theta),
\end{equation}
where the subscript $c$ in $p^{\bm{\alpha}}_c(x_i \mid \theta)$ and $\xi^{\bm{\alpha}}_{c} (\theta \mid \mathbf{x})$ stands for the censoring mechanism.

Censoring offers a practical, low-dimensional alternative to truncating the parameter space (as proposed by \citet{Dimitrakakis:2017:DPB:3122009.3122020}) to achieve an $\epsilon-$DP guarantee, as truncating the parameter space $\Theta$ quickly becomes impractical as the number of parameters grows.

One may imagine a small likelihood value assigned to an extreme observation in the tails.  The identity of such an observation is at a relatively high risk of discovery since its value is isolated from those of other records.  As a result, the absolute value of its log-likelihood, $\abs{f_{\theta}(x_{i})} = \abs{\log{p(x_i \mid \theta)}}$ that determines the $\Delta$ and $\epsilon = 2\Delta$, will be relatively large.  Under the censoring mechanism, the large absolute log-likelihood value will be truncated (in the way shown in Equation~(\ref{eq:censor_like})), which serves to cap the Lipschitz bound.  As compared to a non-censored data distribution over the parameter space, the censored distribution for each $x_{i}$ will be relatively flatter or tempered, removing local features that make detection relatively easier. 

\begin{figure}[t]
  \centering
   \includegraphics[width=0.75\textwidth]{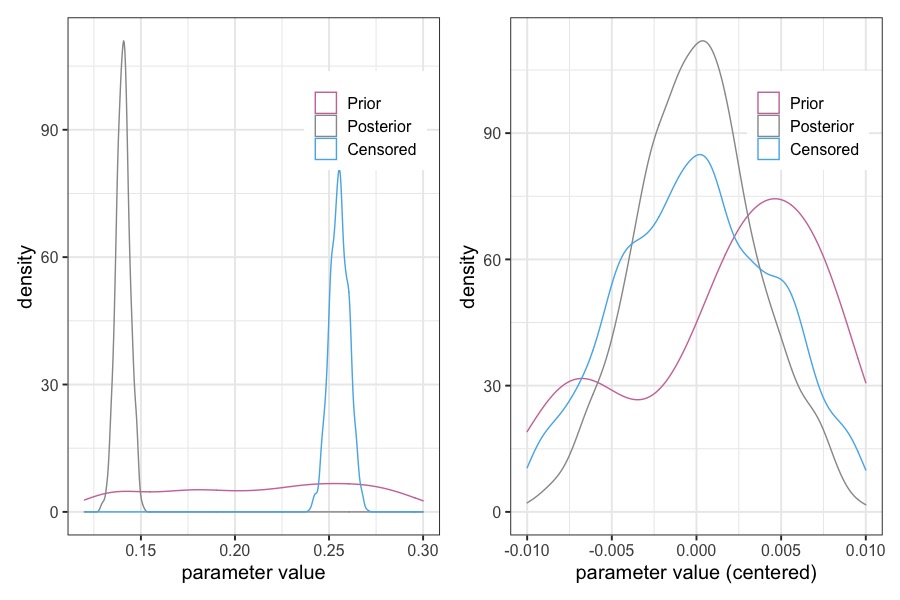}
      \caption{Density plots of of parameter value under the prior, the posterior, and the censoring method in the raw scale (left) and in the centered scale (right).}
      \label{fig:sim:censoring_effects}
\end{figure}

The two panels in Figure \ref{fig:sim:censoring_effects} illustrate the effect of censoring on the resulting parameter posterior distributions. In this illustration, a right-skewed confidential dataset with a long tail of 2000 observations is generated from a $\textrm{Beta}(0.5, 2)$ distribution. Our proposed censored mechanism is implemented at $\epsilon = 1.6$. The right panel of Figure \ref{fig:sim:censoring_effects} centers each density to compare the relative spread of the prior, posterior and censored posterior distributions. Compared to the posterior distribution of a parameter for a given record, the censored distribution is relatively flatter or tempered. This stems from assigning relatively higher (non-absolute) log-likelihood values to parameter values in the tails of the distribution.  The left panel of Figure \ref{fig:sim:censoring_effects} shows that censoring distorts the parameter estimation as compared to the non-censored posterior. These two panels together shed some light on the utility-risk trade-off offered by the censoring mechanism: To achieve strict $\epsilon-$DP guarantee censoring induces the distortion into the censored parameter posterior distribution, which in turn reduces the utility of the resulting synthetic data generated from sampled parameters.

In practice, we do not propose to use a particular dataset for choosing a target $\epsilon$. Instead, one should rely on experience with the class of data (such as data collected on a temporal basis from the same survey instrument). The synthetic data generation process of censored likelihood works in a similar fashion with the $\bm{\alpha}-$weighted pseudo posterior mechanism in Section \ref{sec:methods:weighted} where one uses estimated $\theta^{\ast\ast}$ under the censored likelihood of Equation~(\ref{eq:censor_like}).

In our simulation studies and application in the sequel, we explore both censoring the data likelihood without any downweighting of the records and also formulate an embedding of the $\bm{\alpha}-$weighted pseudo likelihood inside the censoring mechanism.  We label the former mechanism as ``Censor\_uw", which stands for censor unweighted, and the latter as ``Censor\_w", which stands for censor weighted. The Censor\_w construction is a mechanism that would be expected to produce better utility preservation of the real data distribution than the former because censoring is a blunt instrument. The downweighting of likelihood contributions for high-risk records would be expected to invoke censoring less often than under censoring of the likelihood.

We next conduct a Monte Carlo simulation study that generates a collection of size $n$ datasets.  We develop a distribution over the Lipschitz distributions (over the Monte Carlo iterations) to illustrate and compare the contraction of these local Lipschitz bounds. We also evaluate and compare the utility performances of these mechanisms for an equivalent privacy guarantee with PH as a comparison.

\section{Simulation Studies}
\label{sec:sim}

We describe our simulation design in Section \ref{sec:sim:design}. Section \ref{sec:sim:epsilon} presents privacy comparison results by investigating the distributions of local Lipschitz bounds and Section \ref{sec:sim:utility} presents utility comparison results (both global and analysis-specific) with PH as a comparison. We introduce additional techniques to fine tune with downscaling in Section \ref{sec:sim:downscale} and implementation guidance for practitioners using our proposed methods are discussed in Section \ref{sec:sim:guidance}. As will be seen, the Weighted and the Censor\_w synthesizers are the recommended alternatives given their more refined balance of utility-risk trade-off of the resulting synthetic data.

\subsection{Simulation design}
\label{sec:sim:design}

Our simulation studies focus on creating differentially private synthetic data from a confidential dataset generated from a univariate, continuous, and bounded outcome variable. For $r = 1, \cdots, R = 100$, we simulate a local database $\mathbf{x}_r$ of size $n = 2000$ from Beta(0.5, 3) that produces a right-skewed confidential data distribution with many high-risk records in the long tail. For each local database $\mathbf{x}_r$, we fit the Unweighted (i.e., $\alpha_i = 1$ for every record) and obtain its Lipschitz bound, $\Delta_{\bm{\alpha}, \mathbf{x}_r}^{uw}$, and create a synthetic dataset for each of the synthesizers.  We next generate a synthetic dataset for each of the following privacy targets, $\epsilon \in \{5, 4, 3\}$, under each synthesizer.

As discussed in Sections \ref{sec:methods} and \ref{sec:censoring}, the privacy guarantee of each of our four Bayesian data synthesizers is indirectly determined by its calculated Lipschitz bound, after re-estimation of the pseudo posterior. From the implementation point of view, the Weighted-e inputs a target $\epsilon$ and updates or changes some of the weights, $\alpha_{i}$, from the Weighted method to produce a different pseudo posterior estimator. By contrast, Censor\_w also inputs an $\epsilon$ target but does not change the weights $\alpha_{i}$; rather, it may censor the pseudo likelihood contribution for a subset of units.  Censor\_uw works in the same way as Censor\_w, but inputs the unweighted likelihood.

\subsection{Privacy comparison results}
\label{sec:sim:epsilon}

Given that the Weighted and the Weighted-e synthesizers achieve a relaxed aDP privacy guarantee while the Censor\_{uw} and the Censor\_w synthesizers achieve a scrict $\epsilon-$DP guarantee, we expect to see that the distribution of Lipschitz bounds of the Weighted-e is more concentrated at the target $\epsilon$ compared to the Weighted due to its faster contraction rate, whereas the Censor\_w and the Censor\_uw have distributions of Lipschitz bounds strictly at or below the target $\epsilon / 2$.  In what follows, we include the type of privacy guarantee (aDP or DP) in a parenthesis for each synthesizer.

Figure \ref{fig:sim:Lbounds_3es} shows violin plots of the Lipschitz bound distributions over $R = 100$ replicated datasets under our simulation settings. Each panel represents the Lipschitz bound distributions of each synthesizer under one of the three target $\epsilon \in \{5, 4, 3\}$. Note that since the Weighted (aDP) is constructed without taking into account a target $\epsilon$, it is identical across the three panels. The same applies to the Unweighted in a comparison figure included in the Supplementary Materials. Also note that given the $\epsilon = 2 \Delta$ relationship, the scale of the y-axis in Figure \ref{fig:sim:Lbounds_3es}, which shows the Lipscthiz bounds, is half of the target $\epsilon$ value shown at the top of each panel.

As expected, the Weighted-e (aDP) shows a more concentrated distribution of Lipschitz bounds compared to the Weighted (aDP) at every target $\epsilon$. When the distribution of Lipschitz bounds of the Weighted (aDP) is around the target $\epsilon / 2$ (i.e., $\epsilon = 5$ under our simulation settings), the faster concentration gained by the Weighted-e (aDP) is minor. As the target $\epsilon$ decreases, the distribution of Lipschitz bounds gets more concentrated at the target $\epsilon / 2$, particularly obvious when $\epsilon = 3$. In every panel, the Weighted-e (aDP) cannot enforce a strict upper bound of $\epsilon / 2$ of the Lipscthiz bounds over the datasets. In certain cases, as for $\epsilon = 3$ in our simulation, some replicates might have Lipschitz bounds much larger than the target $\epsilon / 2$, as can be seen in the longer upper tail of the Weighted-e (aDP) for $\epsilon = 3$, indicating a lack of control of the privacy guarantee for smaller $\epsilon$ when using the Weighted-e (aDP).

\begin{figure}[t]
  \centering
   \includegraphics[width=0.75\textwidth]{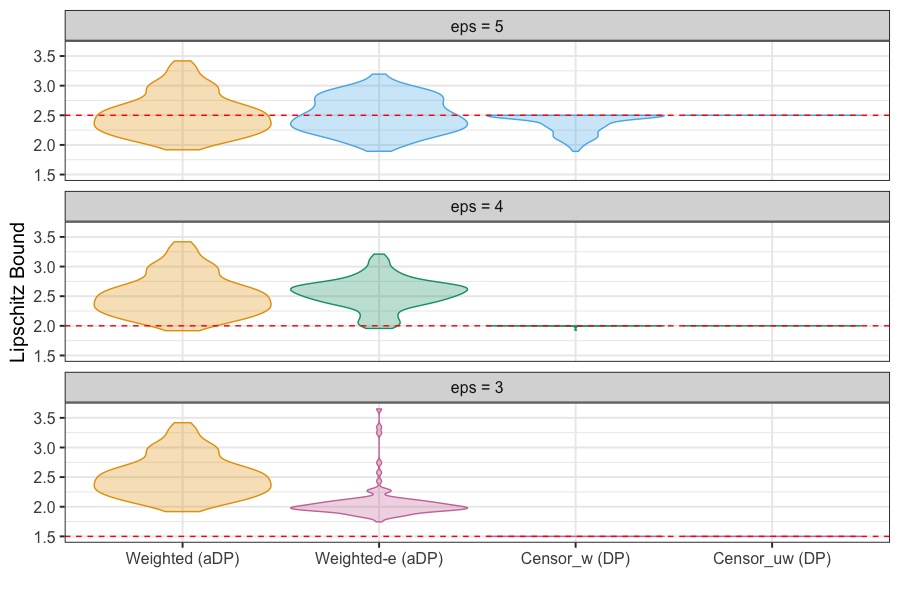}
      \caption{Violin plots of Lipschitz bounds over $R = 100$ replicates under the Weighted (aDP), the Weighted-e (aDP), the Censor\_w (DP), and the Censor\_uw (DP), with $\epsilon$ values of $\{5, 4, 3\}$. A dashed horizontal line at $\epsilon / 2$ is included in each panel.}
      \label{fig:sim:Lbounds_3es}
\end{figure}

The Censor\_w (DP) and the Censor\_uw (DP), by contrast, achieve the strict upper bound of $\epsilon / 2$ for the Lipschitz bounds at every target $\epsilon$, as expected. Among the three pairs of the Censor\_w (DP) and the Censor\_uw (DP), the Censor\_w (DP) shows a clear mass of replicates whose Lipschitz bounds are below $\epsilon / 2$ when $\epsilon = 5$. Recall that the censored likelihood approach directly truncates the pseudo likelihood of each record into [$\exp(-\epsilon / 2), \exp(\epsilon / 2)$]. The difference here is that the Censor\_w (DP) invokes pseudo likelihood censoring of the Weighted (aDP), while the Censor\_uw (DP) does so of the Unweighted. Therefore, the Weighted (aDP) downweights high-risk records which subsequently brings down the Lipshitz bound compared to the Unweighted (excluded in Figure \ref{fig:sim:Lbounds_3es} for ease of reading, the Unweighted shows Lipschitz bounds ranging from around 7.5 to 15, whereas the Weighted (aDP) Lipschitz bounds range from around 2 to 3.5; see the comparison figure in the Supplementary Materials), and the same is happening for the Censor\_w (DP) compared to the Censor\_uw (DP). When $\epsilon = 5$, there may be databases that produce $\epsilon_{\mathbf{x}_{r}} < 5$ under the Weighted (aDP) method such that few if any records have their likelihood values censored in those databases under Censor\_w (DP).  By contrast, since Censor\_uw (DP) embeds the Unweighted posterior synthesizer, every database will require many records to be censored to enforce the $\epsilon/2 = 2.5$ target Lipschitz bound.

\begin{table}[h!]
\centering
\begin{tabular}{ l  l | rrrrrrr }
\hline
 & & Min & Q1 &  Median & Mean & Q3 & Max & sd   \\ \hline 
 $\epsilon = 5$ & Weighted-e (aDP) & 0 & 0 & 0 & 29 & 55 & 166 & 48.80\\
 & Censor\_w (DP) & 0 & 0 & 0 & 110 & 260 & 341 & 129.59\\
 & Censor\_uw (DP) & 196 & 236 & 244 & 247 & 255 & 296 & 17.44 \\ \hline
 $\epsilon = 4$ & Weighted-e (aDP) & 0 & 130 & 219 & 195 & 272 & 364 & 103.48 \\
 & Censor\_w (DP) & 0 & 404 & 425 & 408 & 440 & 549 & 92.56 \\
 & Censor\_uw (DP) & 340 & 398 & 419 & 419 & 435 & 481 & 26.73 \\ \hline
 $\epsilon = 3$ & Weighted-e (aDP) & 400 & 536 & 574 & 573 & 618 & 684 & 60.62 \\
 & Censor\_w (DP) & 655 & 715 & 741 & 741 & 763 & 831 & 35.66 \\
 & Censor\_uw (DP) & 737 & 787 & 806 & 809 & 829 & 888 & 32.85 \\ \hline
\end{tabular}
\caption{Summaries of the number of records (out of $n = 2000$) receiving truncated weight at $\alpha_i$ = 0 in Weighted-e (aDP) and censored likelihood at $\epsilon / 2$ in Censor\_w (DP) and Censor\_uw (DP). The number of Monte Carlo simulations is $R = 100$.}
\label{tab:sim:invokes_count_3es}
\end{table}

To further illustrate the comparison, Table \ref{tab:sim:invokes_count_3es} presents the summaries of the numbers of invocations of truncation in Weighted-e (aDP) and the numbers of invocations of censoring in Censor\_w (DP) and Censor\_uw (DP) for each of the target $\epsilon = \{5, 4, 3\}$ over the $R =100$ Monte Carlo datasets. The summaries include the minimum, Q1, median, mean, Q3, and the maximum of the distribution for invocations of censoring over the $R = 100$ replicates. We also include the standard deviation. Focusing on the three pairs of the Censor\_w (DP) and the Censor\_uw (DP), we can clearly see that for every target $\epsilon$, there are more invocations of censoring in the Censor\_uw (DP) than in the Censor\_w (DP) in terms of the means, resulted from the aforementioned downweighting effects of the Weighted (aDP) in the Censor\_w (DP). Moreover, as $\epsilon$ decreases, the numbers of invocations of censoring increase for both the Censor\_w (DP) and the Censor\_uw (DP). This is expected since a smaller range of [$\exp(-\epsilon / 2), \exp(\epsilon / 2)$] is used in censoring pseudo likelihood as $\epsilon$ decreases, resulting in larger numbers of invocations of censoring. 

Focusing on the number of invocations of truncating $\alpha_i = 0$ for Weighted-e (aDP) in Table \ref{tab:sim:invokes_count_3es}, we observe that same with the censoring, as $\epsilon$ decreases, the number of invocations of truncation increases. The number of invocations of truncation of the Weighted-e (aDP) is overall smaller than the number of invocations of censoring for the Censor\_w (DP).  Yet, despite the reduced number of invocations of censoring over the collection of Monte Carlo datasets, Censor\_w (DP) produces the same censored $\epsilon$ privacy guarantee as does Censor\_uw (DP).





\subsection{Utility comparison results}
\label{sec:sim:utility}

\begin{figure}[t]
  \centering
   \includegraphics[width=0.75\textwidth]{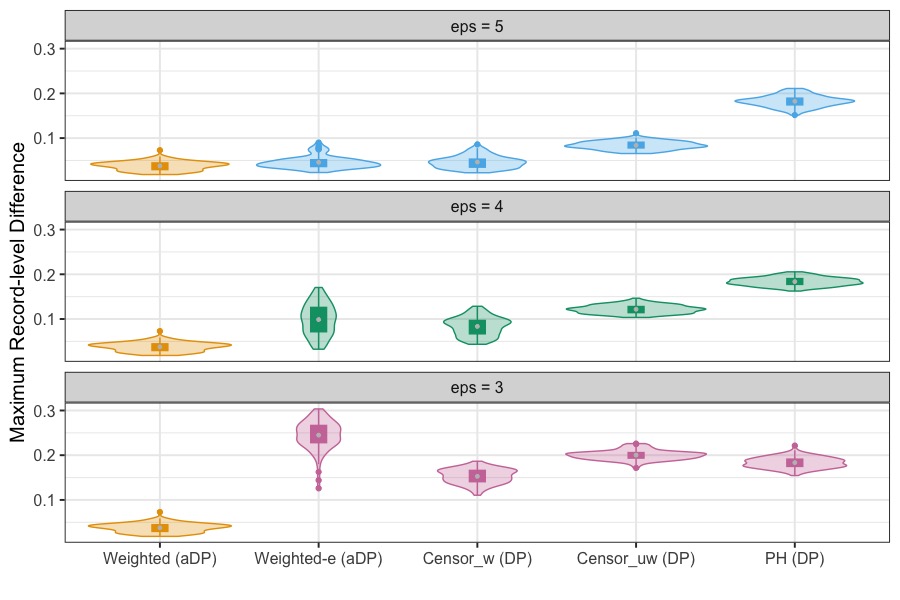}
      \caption{Violin plots of max-ECDF utility over $R = 100$ replicates, for the Weighted (aDP), the Weighted-e (aDP), the Censor\_w (DP), the Censor\_uw (DP), and the PH (DP), with $\epsilon$ values of $\{5, 4, 3\}.$}
      \label{fig:sim:ECDF1}
\end{figure}

\begin{figure}[t]
  \centering
   \includegraphics[width=0.75\textwidth]{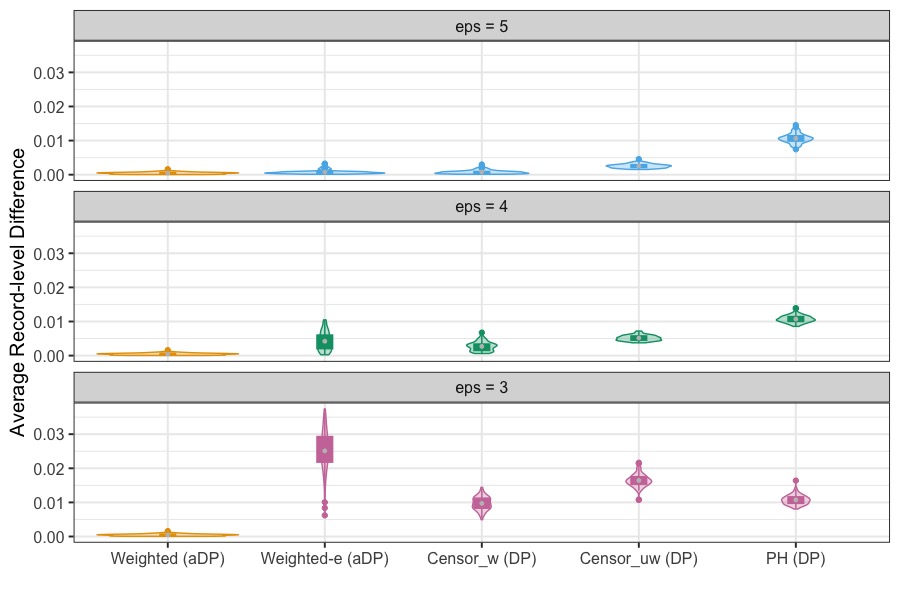}
      \caption{Violin plots of avg-ECDF utility over $R = 100$ replicates, for the Weighted (aDP), the Weighted-e (aDP), the Censor\_w (DP), the Censor\_uw (DP), and the PH (DP), with $\epsilon$ values of $\{5, 4, 3\}.$}
      \label{fig:sim:ECDF2}
\end{figure}

We consider both global utility that describes preservation of the confidential data distribution and analysis-specific utility of the generated synthetic data. We choose the empirical CDF (ECDF) global utility measure, which evaluates whether the distributions for the confidential dataset and the synthetic dataset can be discriminated from each other \citep{Woo2009JPC}. There are two specific measures based on ECDF: 1) the maximum record-level absolute difference (max-ECDF); and 2) the average record-level squared difference (avg-ECDF). Both are calculated as distances between the two ECDFs of the confidential data and the synthetic data. Each measure is non-negative, and a smaller value indicates higher similarity between the two ECDFs, suggesting higher global utility. Violin plots showing the distributions of the max-ECDF for all proposed methods are displayed in Figure \ref{fig:sim:ECDF1}, and those of the avg-ECDF are in Figure \ref{fig:sim:ECDF2}, across $R = 100$ Monte Carlo simulations. As before, each panel represents a target $\epsilon = \{5, 4, 3\}$.

As evident in Figures \ref{fig:sim:ECDF1} and \ref{fig:sim:ECDF2}, the Weighted (aDP) achieves the lowest max-ECDF and avg-ECDF across all synthesizers. These are expected since the Weighted (aDP) is not influenced by a target $\epsilon$ and has the overall highest Lipschitz bounds (i.e., highest risks), as illustrated in Figure \ref{fig:sim:Lbounds_3es}.

Among synthesizers focused on achieving a global privacy target, the Censor\_w (DP) has the highest utility: It expresses the lowest max-ECDF distribution for every target $\epsilon$ in Figure \ref{fig:sim:ECDF1} and also the lowest avg-ECDF distribution for target $\epsilon = \{4, 3\}$ in Figure \ref{fig:sim:ECDF2}.  Censor\_w (DP) embeds the Weighted (aDP) method inside the censoring mechanism, which downweights high-risk likelihood contributions, resulting in less invocations of the censored likelihood.  Although censoring is invoked relatively less often (fixing an $\epsilon$) under embedding of the Weighted (aDP) method, it nevertheless comes equipped with an $\epsilon-$DP guarantee because the censoring procedure is applied to all datasets.  The Weighted-e (aDP) method, by contrast, seems to have an even more concentrated and lower distribution when target $\epsilon = 5$. However, as $\epsilon$ decreases, the Weighted-e (aDP)'s performance starts to deteriorate rapidly, which indicates that it lacks not only the control of the privacy guarantee (shown in Figure \ref{fig:sim:Lbounds_3es}) but also the control of utility preservation. The Censor\_uw (DP) performs consistently worse than the Censor\_w (DP) at every target $\epsilon$ value, an expected result since the Censor\_uw (DP) invokes censoring for a larger number of records than the Censor\_w (DP) (shown in Table \ref{tab:sim:invokes_count_3es}) which should result in lower utility. The PH (DP), a non-Bayesian data synthesizer included for comparison, clearly performs worse than the almost all proposed Bayesian synthesizers. Overall, all synthesizers (except for the Weighted (aDP) which is not affected by the target $\epsilon$ value) have lower utility as $\epsilon$ decreases.

\begin{figure}[t]
  \centering
   \includegraphics[width=0.75\textwidth]{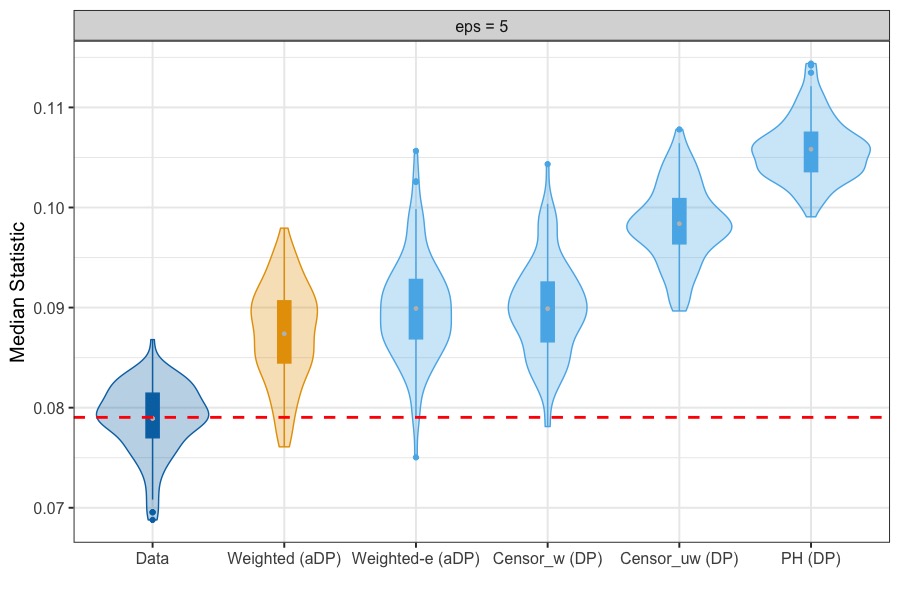}
      \caption{Violin plots of median over $R = 100$ replicates for the Weighted (aDP), the Weighted-e (aDP), the Censor\_w (DP), the Censor\_uw (DP), and the PH (DP), at $\epsilon = 5$. A dashed horizontal line at the analytical median from Beta(0.5, 3) is included.}
      \label{fig:sim:median_e5}
\end{figure}

Figures \ref{fig:sim:median_e5} displays violin plots of the distributions of the median statistic across $R = 100$ Monte Carlo simulations for $\epsilon = 5$. (Results of the 15th and the 90th quantile statistics and all three statistics for $\epsilon = \{4, 3\}$ tell a similar story and are in the Supplementary Materials.) The distributions from the confidential data are included and labeled as ``Data" for comparison. A dashed horizontal line marking the analytical value from Beta(0.5, 3), the generating distribution, is included in each plot. The closer the resulting violin plot from a data synthesizer is to the one for Data and the dashed horizontal line, the higher the utility is for that statistic.

As with the global utility evaluation, the Weighted (aDP) has overall the highest utility, at the price of achieving an aDP guarantee. Among the remaining synthesizers, the Censor\_w (DP) once again shows the highest utility performance for the median statistic. 
The Weighted-e (aDP) overall has better utility performance on these analysis-specific utility measures as compared to the global utility measures discussed earlier. The Censor\_uw (DP) performs worse than the Censor\_w (DP) in all statistics because censoring is invoked on more data records in every dataset. 

\subsection{Fine tuning with downscaling}
\label{sec:sim:downscale}

As Figures \ref{fig:sim:ECDF1} and \ref{fig:sim:ECDF2} of global utility have shown, as $\epsilon$ decreases from 5 to 4 or 3, the utility performances of Censor\_w (DP) deteriorates because more invocations of censoring across records occur as $\epsilon$ decreases. Further scaling and shifting of the weights $\bm{\alpha}$ in the Weighted (aDP) method may be used in order to achieve a target privacy guarantee for any confidential dataset; for example, by setting $(c_1, c_2)$ to $\tilde{\bm{\alpha}} \in [0, 1]$ (as in $\tilde{\alpha}_{i} = c_1 \times \alpha_{i} + c_2$). Instead of using $c_1 = 1$ as we have done in Sections \ref{sec:sim:epsilon} and \ref{sec:sim:utility}, we use a scaling factor $c_1 < 1$ that downscales the privacy weights $\bm{\alpha}$.  The weight scaling allows us to roughly target a specific $\epsilon$ and censoring is then used to make that target global.  This procedure reduces the number of invocations of censoring while preserving a strict $\epsilon-$DP guarantee.

\begin{figure}[t]
  \centering
   \includegraphics[width=0.75\textwidth]{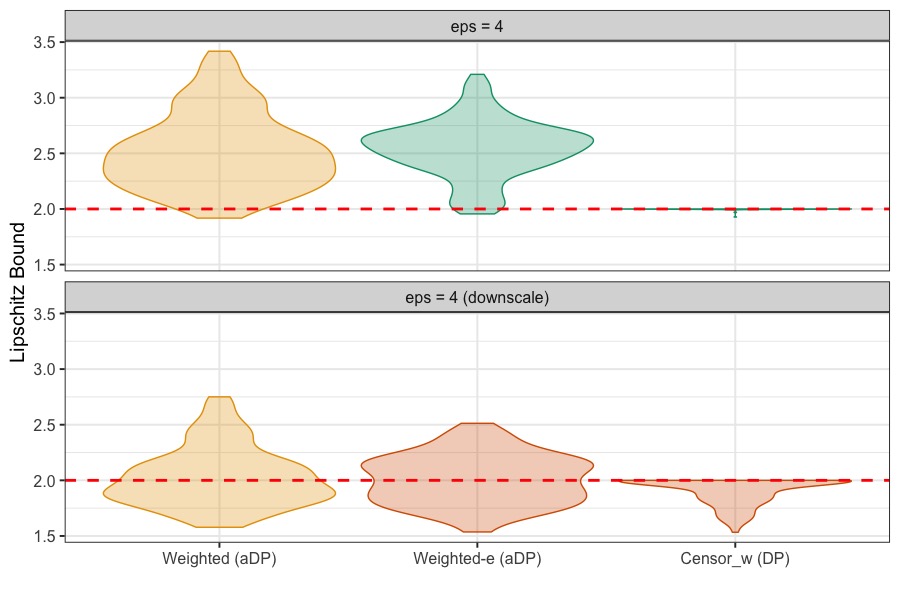}
     \caption{Violin plots of Lipschitz bounds over $R = 100$ replicates under the Weighted (aDP), the Weighted-e (aDP), and the Censor\_w (DP), at $\epsilon = 4$, without downscaling (top) and with downscaling (bottom). A dashed horizontal line at $\epsilon / 2$ is included.}
      \label{fig:sim:Lbounds_e4}
\end{figure}

\begin{figure}[t]
  \centering
   \includegraphics[width=0.75\textwidth]{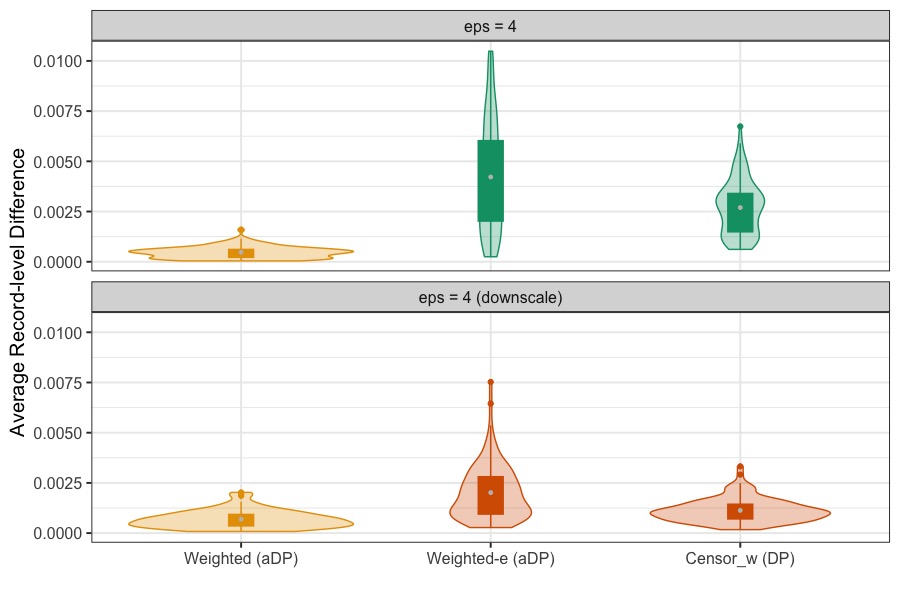}
      \caption{Violin plots of avg-ECDF utility over $R = 100$ replicates, for the Weighted (aDP), the Weighted-e (aDP), and the Censor\_w (DP) at $\epsilon = 4$, without downscaling (top) and with downscaling (bottom).}
      \label{fig:sim:ECDF4}
\end{figure}

\begin{figure}[t]
  \centering
   \includegraphics[width=0.75\textwidth]{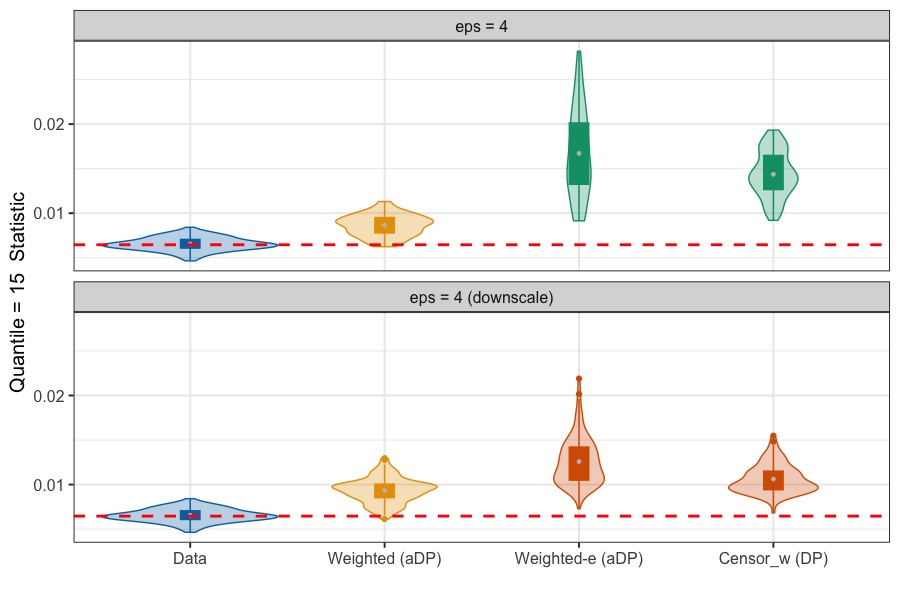}
      \caption{Violin plots of the 15th quantile over $R = 100$ replicates for the Weighted (aDP), the Weighted-e (aDP), and the Censor\_w (DP) at $\epsilon = 4$, without downscaling (top) and with downscaling (bottom). A dashed horizontal line at the analytical 15th quantile from Beta(0.5, 3) is included.}
      \label{fig:sim:q15_e4}
\end{figure}

We focus on $\epsilon = 4$ for illustration. We set $c_1 = 0.8$ for the results that follow. Since only the Weighted-e (aDP) and the Censor\_w (DP) depend on the Weighted (aDP) method whereas Censor\_uw (DP) embeds the Unweighted method, we expect to see effects of downscaling on the Weighted (aDP), the Weighted-e (aDP), and the Censor\_w (DP) synthesizers. Therefore, we focus our comparison on these three synthesizers. 

To evaluate the privacy guarantee and utility performances without and with downscaling, we generate $R = 100$ replicate datasets in a Monte Carlo simulation study. Figure \ref{fig:sim:Lbounds_e4} shows 
with downscaling, the Lipschitz bounds are lower and more concentrated than those without downscaling, indicating a better control of the privacy guarantee for these three synthesizers. Figure \ref{fig:sim:ECDF4} illustrates the effects of downscaling on the avg-ECDF based global utility measure (results of the max-ECDF are included in the Supplementary Materials). 
As can be seen, the violin plots of the Weighted (aDP), the Weighted-e (aDP), and the Censor\_w (DP) are all shifted to lower values under downscaling, indicating higher utility. 
Moreover, as shown in Figure \ref{fig:sim:q15_e4} for the 15th quantile statistic utility, downscaling also improves this and other analysis-specific utility measures (available in the Supplementary Materials).

\subsection{Implementation guidance}
\label{sec:sim:guidance}
We next describe a recommended procedure to achieve some target $\epsilon$ for a collection of time-indexed datasets (such as the monthly release of total employment for each state) that uses censoring to acheive a strict $\epsilon-$DP guarantee.  We note that our implementation suggestions apply when using any reference dataset with relaxed aDP guarantees, not necessarily those that rely on regularly scheduled releases and temporal autocorrelation.

Using the current period database to set the privacy $\epsilon$ for any mechanism can risk leaking a small amount of information about the overall sensitivity, $\Delta$, of the database under a strict $\epsilon-$DP guarantee.    Under the pseudo posterior mechanism equipped with a relaxed, asymptotic DP guarantee, however, the risk of disclosure about the sensitivity of the database is very low.  The reason for the low risk is because the record-indexed weights under the pseudo posterior are set to be inversely proportional to their by-record sensitivities such that the resulting mechanism overall sensitivity (and thus local $\epsilon$) after weighting is nearly unrelated to the underlying sensitivity of the local database.    We see this by the increased concentration of the distribution of by-record sensitivities after weighting from before weighting for the pseudo posterior mechanism \citep{SavitskyWilliamsHu2020ppm}.   (We remind the reader that the resulting weights are not published). 

We recommend utilizing the censoring mechanism with the pseudo posterior to achieve a strict $\epsilon-$DP guarantee, though we propose to use the sensitivity and utility measures of a \textit{historical} (rather than the current) database to set the global $\epsilon$ under the censoring mechanism in an abundance of caution.    In practice, the owner of the confidential data (such as a government statistical agency) publishes a database of size $n$ on a regular, periodic basis (e.g., the publication of monthly employment and unemployment by the U.S. Bureau of Labor Statistics).   

In the first period of the year, we recommend that the owner set the privacy guarantee $\epsilon$ by first constructing $\bm{\alpha} = (\alpha_{1},\ldots,\alpha_{n})$ as outlined in Section~\ref{sec:methods:weighted}, and re-estimating the model under the pseudo posterior mechanism to compute the associated $\Delta_{\alpha}$.  This gives an asymptotic aDP guarantee of $2\Delta_{\alpha}$.  

Next, we recommend to shift and scale the weights to an $\tilde{\bm{\alpha}}$ that produces a target $\epsilon$ on the first period dataset under an aDP guarantee.  This may take repeated model re-estimations to evaluate the utility for each shifting and scaling of the weights to assess the risk-utility trade-off needed to pick the target $\tilde{\bm{\alpha}}$.    

In future months, the \emph{same} $\epsilon$ would be used to censor the pseudo likelihood contributions with the weights scaled and shifted by the same $(c_{1},c_{2})$ as used in the first month.  In this way, the many computations used to set $(c_{1},c_{2},\epsilon)$ are performed only once.  These settings will likely result in fewer invocations of censoring to produce synthetic data in future months under an $\epsilon-$DP guarantee such that the resulting utility of the synthetic data would be expected to be similar to the first month.

\section{Application to Survey of Doctoral Recipients}
\label{sec:app}

In this application section, we apply the Weighted (aDP), the Weighted-e (aDP), the Censor\_w (DP), the Censor\_uw (DP), and the PH (DP) synthesizers to a sample of the Survey of Doctoral Recipients. Our sample comes from the public use file published for 2017. It contains information on salary, gender, age (grouped in 5-year intervals), and the number of working weeks for $n = 1601$ survey respondents who have positions at a 4-year college or university in the field of mathematics and statistics. Our Unweighted model is a beta regression, where the outcome variable salary is scaled into [0, 1] and gender (categorical with 2 levels), age (categorical with 9 levels), and weeks (numerical ranging from 2 weeks to 52 weeks) as publicly available predictors. The target privacy budget is $\epsilon = 5$, which is used in our implementation of the perturbed histogram synthesizer for comparison. For every synthesizer, one synthetic dataset is simulated.

\begin{figure}[t]
  \centering
   \includegraphics[width=0.75\textwidth]{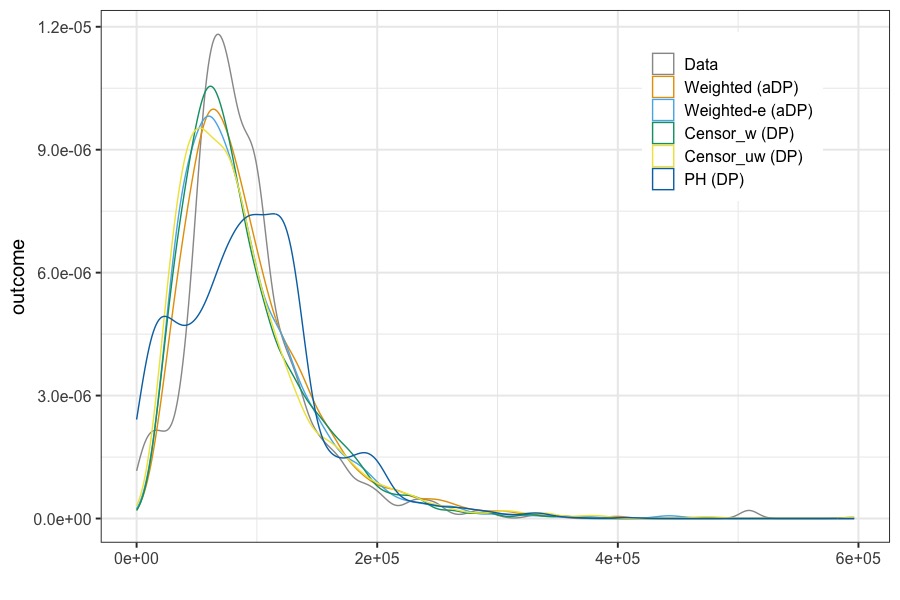}
   \caption{Density plots of the confidential data sample (Data) and the synthetic data samples from the Weighted (aDP), the Weighted-e (aDP), the Censor\_w (DP), the Censor\_uw (DP), and the PH (DP) synthesizers at $\epsilon = 5$. }
   \label{fig:app:density}
\end{figure}

Figure \ref{fig:app:density} displays density plots of the confidential data sample and those from the synthetic datasets simulated from the five synthesizers. The PH (DP) evidently has the worst utility performance in this visual check. Among the four Bayesian synthesizers, the Censor\_w (DP) seems to capture the main feature of the confidential data the best, with the Weighted (aDP) coming in second, while the Weighted-e (aDP) and the Censor\_uw (DP) come in third and fourth, respectively.

Table \ref{tab:app_utility1} presents detailed privacy and utility results of our synthesizers. In terms of privacy, we include the calculated Lipschitz bound of the Weighted (aDP), the Weighted-e (aDP), the Censor\_w (DP), and the Censor\_uw (DP) and their corresponding $\epsilon$ values. Note that Data and the PH (DP) have no Lipschitz bounds and therefore coded as NA, and Data has no privacy guarantee which is coded as NA as well. As we know from Sections \ref{sec:methods} and \ref{sec:censoring} and the simulation study results in Section \ref{sec:sim:epsilon}, the Censor\_w (DP) and the Censor\_uw (DP) have strict Lipschitz bound at $\epsilon / 2$, whereas the Weighted (aDP) and the Weighted-e (aDP) only contract to $\epsilon / 2$ but cannot guarantee $\epsilon / 2$ (with the Weighted-e (aDP) contracting faster than the Weighted (aDP)). As expected, in our application, the Censor\_w (DP) and the Censor\_uw (DP) achieve the target $\epsilon = 5$ whereas the Weighted (aDP) and the Weighted-e (aDP) have privacy guarantee 5.24 and 5.22 respectively, both exceed the target privacy budget.

\begin{table}[ht]
\centering
\begin{tabular}{ l | r | rrrrrr }
\hline
 & Data & Weighted & Weighted-e & Censor\_w & Censor\_uw & PH   \\ 
 & & (aDP) &  (aDP) &  (DP) &  (DP) &  (DP) \\ \hline 
Lipschitz & NA & 2.62 & 2.61 & 2.50 & 2.50 & NA \\ 
Privacy $\epsilon$ & NA & 5.24 & 5.22 & 5.00 & 5.00 & 5.00 \\ \hline
max-ECDF & NA &  {\bf 0.0656}   &     0.1020 &   {\underline{0.0968}}    & 0.1350 & 0.1310\\
avg-ECDF & NA &  {\bf 0.0011} &     {\underline{0.0025}} &  0.0026 &  0.0039 & 0.0057 \\ 
Mean & 91019 &  {\underline{92994}} &  {\bf 89525} &  88581  & 88840 &  93654 \\
Median & 80000  & {\bf 80135} &  {\underline{76451}} &  75642  & 75211  & 91180 \\
15th Q & 51000 &  {\bf 44303}  & 40574  & {\underline{41142}}   & 38107 &  30108 \\
90th Q & 150000 &  162351 &   {\bf 158037} &   {\underline{158426}} &   162686 &   163872 \\ \hline
\end{tabular}
\caption{Privacy and utility results of the Weighted (aDP), the Weighted-e (aDP), the Censor\_w (DP), the Censor\_uw (DP), and the PH (DP) at $\epsilon = 5$. In the utility rows, the best performing synthesizer is in bold and the second best is underlined.}
\label{tab:app_utility1}
\end{table}

Table \ref{tab:app_utility1} also includes global utility and analysis-specific utility results for all synthesizers. For the two global utility measures, max-ECDF and avg-ECDF, the smaller the value, the higher the utility. (Data has no such measures and therefore coded as NA.) For the four analysis-specific utility measures, the mean, the median, the 15th quantile, and the 90th quantile statistics, the closer they are to the value from the Data, the higher the utility. For ease of reading, the value of the best performing synthesizer among the five for each statistic is in bold, and the second best is underlined. 

Undoubtedly, the Weighted (aDP) has the highest utility performance, and the Weighted-e (aDP) comes second while the Censor\_w (DP) comes third. The inferior results of the Censor\_uw (DP) and the PH (DP) are expected: The Censor\_uw (DP) embeds the Unweighted synthesizer into censoring, which results in the invocation of censoring for more records than Censor\_w (DP).  The PH (DP) does not take advantage of the benefits from modeling and the useful predictor variables available as public information.  Although the Weighted-e (aDP) performs slightly better than the Censor\_w (DP) in terms of utility in our application, such improvement cannot offset the fact that it provides an asymptotic privacy guarantee. 

\section{Concluding Remarks}
\label{sec:conclusion}

In this paper, we review the pseudo posterior mechanism from \citep{SavitskyWilliamsHu2020ppm} which achieves an asymptotic DP guarantee. To provide a stronger, non-asymptotic DP guarantee, we propose the censoring mechanism that censors the pseudo likelihood of every record within $[\exp(-\epsilon / 2), \exp(\epsilon / 2)]$. This new mechanism truncates large absolute log-likelihood values, which serves to cap the Lipschitz bound and therefore achieving an $\epsilon-$DP guarantee. It offers a practical, low-dimensional alternative to truncating the parameter space.

Our simulation studies on [0, 1] bounded univariate data under repeated sampling demonstrate the superior utility preservation performance of the pseudo posterior mechanism at the cost of an asymptotic DP guarantee, on the one hand. The censoring mechanism that embeds the pseudo posterior mechanism provides a stronger, non-asymptotic DP guarantee at the price of slightly reduced utility performance, on the other hand. We recommend these two alternatives for data disseminators who depending on their priority, can choose the preferred strategy between these two. 

One interesting line of future work is embedding other weighted synthesizers in censoring to provide a non-asymptotic DP guarantee. Another is to investigate, in an applied setting, how similar do the data distributions at two points in time need to be for their utilities to be similar under the same scaling, shifting, and privacy loss parameters. 

\section*{Supplementary Materials}

Supplementary Materials include a detailed review of the perturbed histogram synthesizer, the Stan script for the censoring method, and additional privacy and utility comparison results from Section 4.
\par
\section*{Acknowledgements}

This research was supported, in part, by the National
Science Foundation (NSF), National Center for Science and Engineering Statistics (NCSES) by the Oak Ridge Institute for
Science and Education (ORISE) for the Department of Energy (DOE). ORISE is managed by Oak Ridge Associated Universities
(ORAU) under DOE contract number DE-SC0014664. The authors also wish to thank Claire Bowen, a Lead Data Scientist of Privacy and Data Security at The Urban Institute, for her advice on our implementation of the perturbed histogram synthesizer. All opinions expressed in this paper are the authors' and do not necessarily reflect the policies and views of NSF, BLS, DOE, ORAU, or ORISE.
\par


\bibhang=1.7pc
\bibsep=2pt
\fontsize{9}{14pt plus.8pt minus .6pt}\selectfont
\renewcommand\bibname{\large \bf References}
\expandafter\ifx\csname
natexlab\endcsname\relax\def\natexlab#1{#1}\fi
\expandafter\ifx\csname url\endcsname\relax
  \def\url#1{\texttt{#1}}\fi
\expandafter\ifx\csname urlprefix\endcsname\relax\def\urlprefix{URL}\fi


\bibliographystyle{ba}
\bibliography{DPbib}

\vskip .65cm
\noindent
Vassar College, Box 27, 124 Raymond Ave, Poughkeepsie, NY 12604 
\vskip 2pt
\noindent
E-mail: jihu@vassar.edu
\vskip 2pt

\noindent
RTI International, 3040 East Cornwallis Road, Research Triangle Park, NC 27709
\vskip 2pt
\noindent
E-mail: mrwilliams@rti.org

\noindent
U.S. Bureau of Labor Statistics, Office of Survey Methods Research, Suite 5930, 2 Massachusetts Ave NE Washington, DC 20212
\vskip 2pt
\noindent
E-mail: Savitsky.Terrance@bls.gov


\section*{Supp 1: Detailed review of the perturbed histogram method in Section 2.3}

We present the commonly-used perturbed histogram mechanism for simulating synthetic microdata that achieves $\epsilon-$DP guarantee \citep{Dwork:2006:CNS:2180286.2180305, WassermanZhou2010} as a comparison. Under the required strong assumption of a \emph{bounded} and continuous variable, one first discretizes it into a histogram with a selected number of bins. One induces a formal $\epsilon-$DP privacy guarantee into the histogram by adding Laplace noise. The $\epsilon-$DP guarantee is only global to the extent that one assumes the data space of datasets of size $n$, $\mathcal{X}^{n}$ is absolutely bounded, which is highly unlikely in practice.  Finally, one simulates microdata from the private histogram under $\epsilon-$DP, which is a post-processing step in a similar fashion as generating synthetic data under the pseudo posterior mechanism (given the privacy protected parameter draws) reviewed in Section 2.1. 

We describe the perturbed histogram synthesizer for univariate data, $\mathbf{x}$, of size $n$, with a bound of size $L$.  We select the number of bins, $m$, that we use to partition $\mathbf{x}$ into $m$ bins, $\{B_1, \cdots, B_m\}$, where each bin, $B_j$, is of length $L / m$. The choice of $m$ should be independent of the data $\mathbf{x}$; e.g., $m = \ln(n)$ or $m = \sqrt{n}$. Let $C_j = \sum_{i=1}^{n} I(x_i \in B_j)$, where $I(\cdot)$ is the indicator function; i.e., $C_j$ is the number of observations in bin $B_j$. 

To privatize the resulting histogram, let $D_j = C_j + \eta_j$, where $\eta_1, \cdots, \eta_m$ $\overset{i.i.d.}{\sim} \textrm{Laplace}(0, 2 / \epsilon)$. In other words, each bin count $C_j$ has added noise from a Laplace distribution with mean 0 and scale $2 / \epsilon$, where $\epsilon$ is the targeted privacy budget and $2$ is the global sensitivity of a histogram (to reflect the move of a unit from one bin to another). This noise addition process guarantees $\epsilon-$DP for $\bm D = (D_1, \cdots, D_m)$ \citep{Dwork:2006:CNS:2180286.2180305, WassermanZhou2010}. 

Finally, to create synthetic microdata from private $\bm D$, define $\tilde{D}_j = \textrm{max}(D_j, 0)$. Calculate $\hat{q}_j = \tilde{D}_j / \sum_s \tilde{D}_s$, which is the probability of membership in each privatized bin $B_j$. To simulate a synthetic microdata value for record $i$, we first take a multinomial draw under probabibilities $(\hat{q}_1, \cdots, \hat{q}_m)$, resulting in a bin indicator $b_i \in (1, \cdots, m)$. Next, given the sampled bin indicator $b_i$, we take a uniform draw from that bin to generate synthetic value $x_i^*$ for record $i$. We repeat this process for all records, obtaining a synthetic dataset $\mathbf{x}^*$ from the perturbed histogram synthesizer. 

\section*{Supp 2: Stan script for the censoring method for the beta synthesizer used in Section 4}

Below we include the Stan script for our censoring method for the beta synthesizer used in our simulation study in Section 4. Note that the value \texttt{M} in the script is set as $M = \epsilon / 2$.


\begin{verbatim}

functions{

  real betawt_lpdf(vector outcome, real beta1, real beta2, vector alpha, int n, real M)
  {
    real check_term;
    real update_term;
    check_term  = 0.0;
    for( i in 1:n )
    {
        update_term = alpha[i] * beta_lpdf(outcome[i] | beta1, beta2);
	      check_term    = check_term + fmax(fmin(update_term, M), -M);
    }
    return check_term;
  }

  real betawt_i_lpdf(real outcome_i, real beta1, real beta2, real alpha_i, real M){
    real check_term;
    real update_term;
	    update_term   = alpha_i * beta_lpdf(outcome_i | beta1, beta2);
	    check_term    = fmax(fmin(update_term, M), -M);
    return check_term;
  }
  
  real betawt_i_noM_lpdf(real outcome_i, real beta1, real beta2, real alpha_i){
    real check_term;
	    check_term    = alpha_i * beta_lpdf(outcome_i | beta1, beta2);
    return check_term;
  }
  
  
} /* end function{} block */

data {
    int<lower=1> n; // number of observations
    vector[n] outcome; // Response variable
    vector<lower=0>[n] alpha; // observation-indexed (privacy) weights
    real<lower=0> M; //censoring threshold
}

parameters{
  real<lower=0,upper=1> phi;
  real<lower=0.1> lambda;
}

transformed parameters{
  real<lower=0> beta1 = lambda * phi;
  real<lower=0> beta2 = lambda * (1 - phi);
}

model{
  phi ~ beta(1, 1); // uniform on phi, could drop
  lambda ~ pareto(0.1, 1.5);
  
  target        += betawt_lpdf(outcome | beta1, beta2, alpha, n, M);
} /* end model{} block */

generated quantities{
  vector[n] log_lik;
  vector[n] log_lik_noM;
  
  for (i in 1:n) {
	log_lik[i] = betawt_i_lpdf(outcome[i] | beta1, beta2, alpha[i], M);
	log_lik_noM[i] = betawt_i_noM_lpdf(outcome[i] | beta1, beta2, alpha[i]);
	}
}
%\end{lstlisting}
%\end{Verbatim}
\end{verbatim}

\section*{Supp 3: Additional privacy comparison results from Section 4.2}

\begin{figure}[H]
  \centering
   \includegraphics[width=0.95\textwidth]{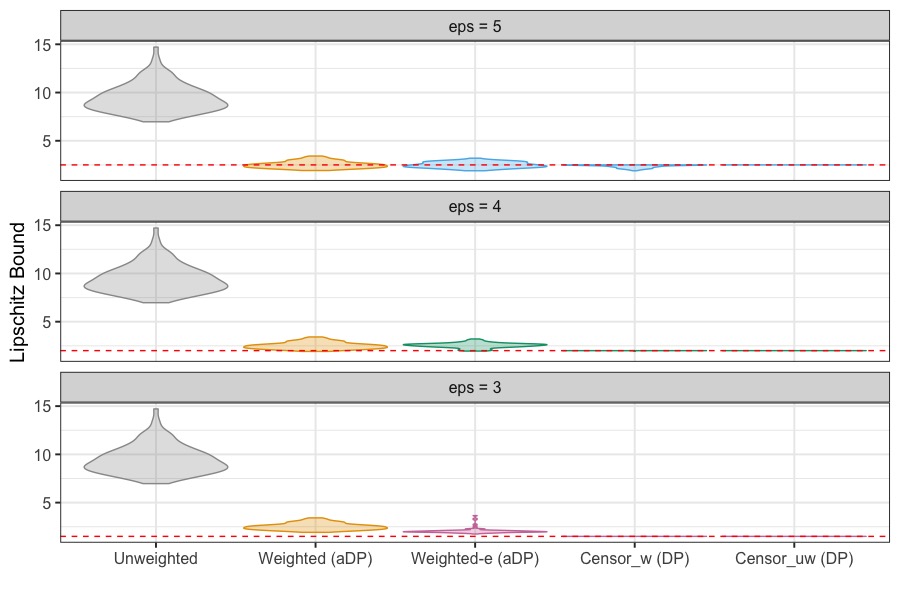}
      \caption{Violin plots of Lipschitz bounds over $R = 100$ replicates under the Unweighted, the Weighted (aDP), the Weighted-e (aDP), the Censor\_w (DP), and the Censor\_uw (aDP), with $\epsilon$ values of $\{5, 4, 3\}$. A dashed horizontal line at $\epsilon / 2$ is included in each panel.}
      \label{fig:sim:Lbounds_3es}
\end{figure}

\section{Additional utility comparison results from Section 4.3}

\begin{figure}[H]
  \centering
   \includegraphics[width=0.8\textwidth]{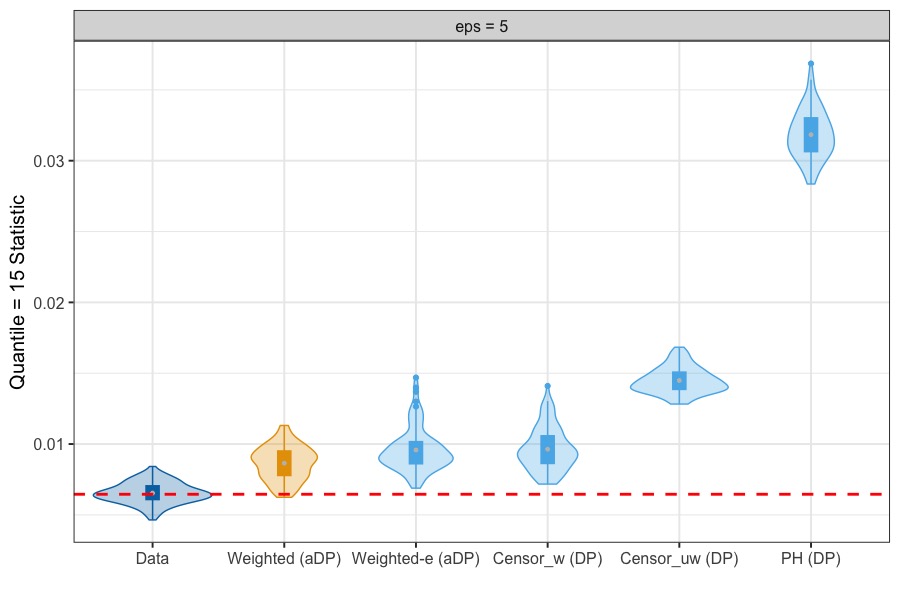}
      \caption{Violin plots of the 15th quantile over $R = 100$ replicates for the Weighted (aDP), the Weighted-e (aDP), the Censor\_w (DP), the Censor\_uw (DP), and the PH (DP), at $\epsilon = 5$. A dashed horizontal line at the analytical 15th quantile from Beta(0.5, 3) is included.}
      \label{fig:sim:Q15_e5}
\end{figure}

\begin{figure}[H]
  \centering
   \includegraphics[width=0.8\textwidth]{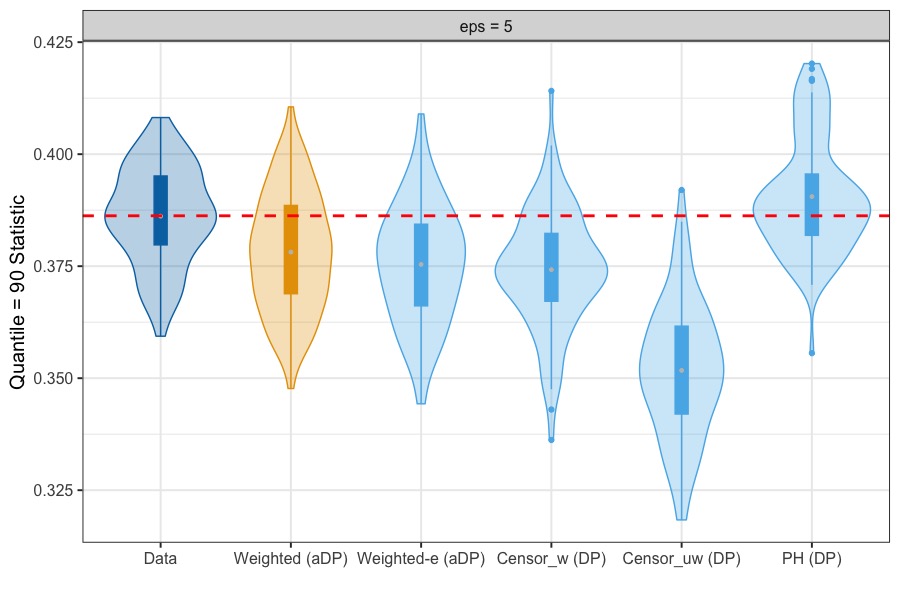}
      \caption{Violin plots of the 90th quantile over $R = 100$ replicates for the Weighted (aDP), the Weighted-e (aDP), the Censor\_w (DP), the Censor\_uw (DP), and the PH (DP), at $\epsilon = 5$. A dashed horizontal line at the analytical 90th quantile from Beta(0.5, 3) is included.}
      \label{fig:sim:Q90_e5}
\end{figure}

\begin{figure}[H]
  \centering
   \includegraphics[width=0.8\textwidth]{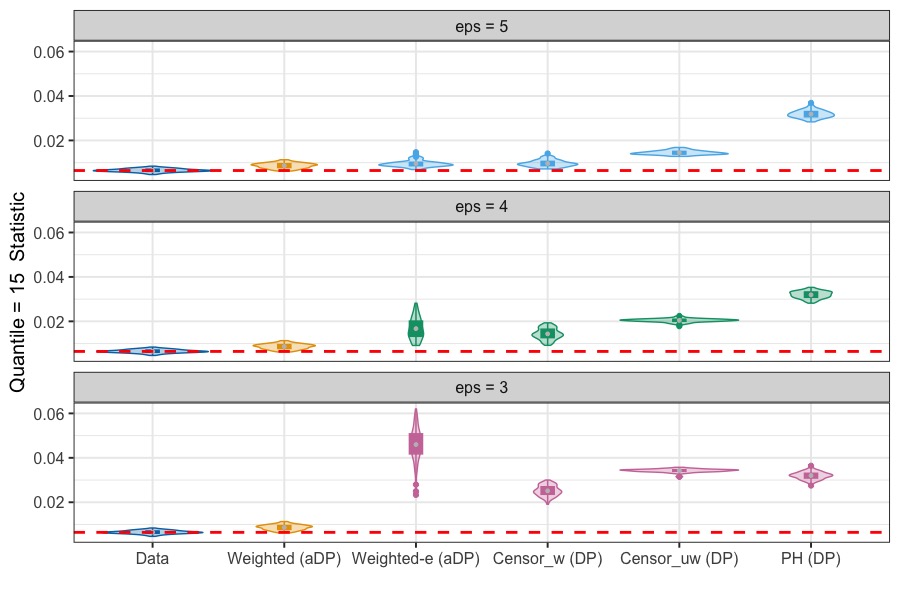}
      \caption{Violin plots of the 15th quantile over $R = 100$ replicates, for the Weighted (aDP), the Weighted-e (aDP), the Censor\_w (DP), the Censor\_uw (DP), and the PH (DP), with $\epsilon$ values of $\{5, 4, 3\}.$ A dashed horizontal line at the analytical 15h quantile from Beta(0.5, 3) is included in each panel.}
\end{figure}

\begin{figure}[H]
  \centering
   \includegraphics[width=0.8\textwidth]{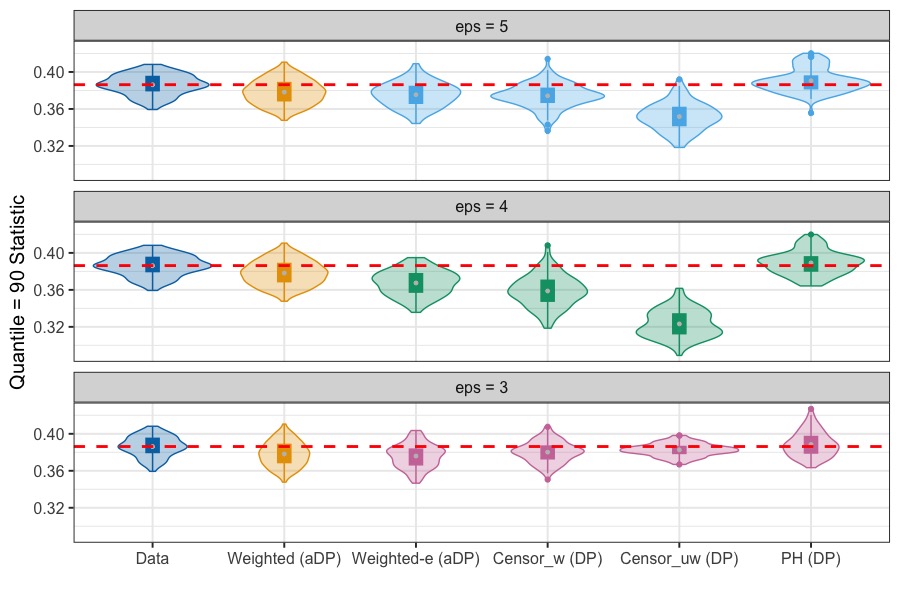}
      \caption{Violin plots of the 90th quantile over $R = 100$ replicates, for the Weighted (aDP), the Weighted-e (aDP), the Censor\_w (DP), the Censor\_uw (DP), and the PH (DP), with $\epsilon$ values of $\{5, 4, 3\}.$ A dashed horizontal line at the analytical 90th quantile from Beta(0.5, 3) is included in each panel.}
\end{figure}

\begin{figure}[H]
  \centering
   \includegraphics[width=0.8\textwidth]{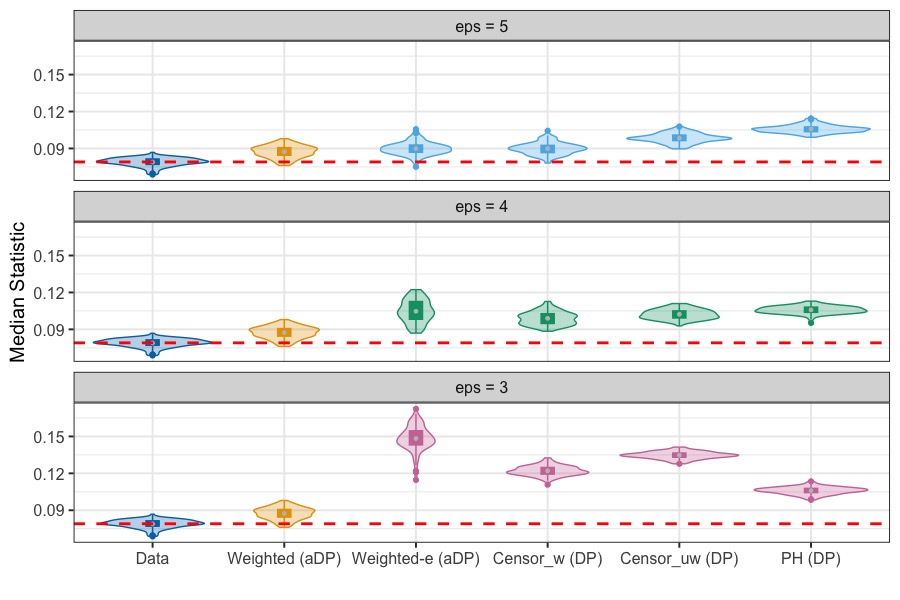}
     \caption{Violin plots of the median over $R = 100$ replicates, for the Weighted (aDP), the Weighted-e (aDP), the Censor\_w (DP), the Censor\_uw (DP), and the PH (DP), with $\epsilon$ values of $\{5, 4, 3\}.$ A dashed horizontal line at the analytical median from Beta(0.5, 3) is included in each panel.}
\end{figure}

\section*{Supp 4: Privacy and additional utility comparison results from Section 4.4}

\begin{table}[H]
\centering
\begin{tabular}{ l  l | rrrrrrr }
\hline
 & & Min & Q1 &  Median & Mean & Q3 & Max & sd   \\ \hline 
 $\epsilon = 4$ & Weighted-e (aDP) & 0 & 130 & 219 & 195 & 272 & 364 & 103.48 \\
 & Censor\_w (DP) & 0 & 404 & 425 & 408 & 440 & 549 & 92.56 \\ \hline
 $\epsilon = 4$  & Weighted-e (aDP) & 0 & 0 & 93 & 91 & 165 & 269 & 86.49 \\
 (downscale) & Censor\_w  (DP) & 0 & 0 & 0 & 144 & 328 & 413 & 165.66 \\ \hline
\end{tabular}
\caption{Summaries of the number of records (out of $n = 2000$) receiving truncated weight at $\alpha_i$ = 0 in Weighted-e (aDP) and censored likelihood at $\epsilon / 2$ in Censor\_w (DP). The number of Monte Carlo simulations is $R = 100$.}
\label{tab:sim:invokes_count_e4}
\end{table}
\begin{figure}[H]
  \centering
   \includegraphics[width=0.8\textwidth]{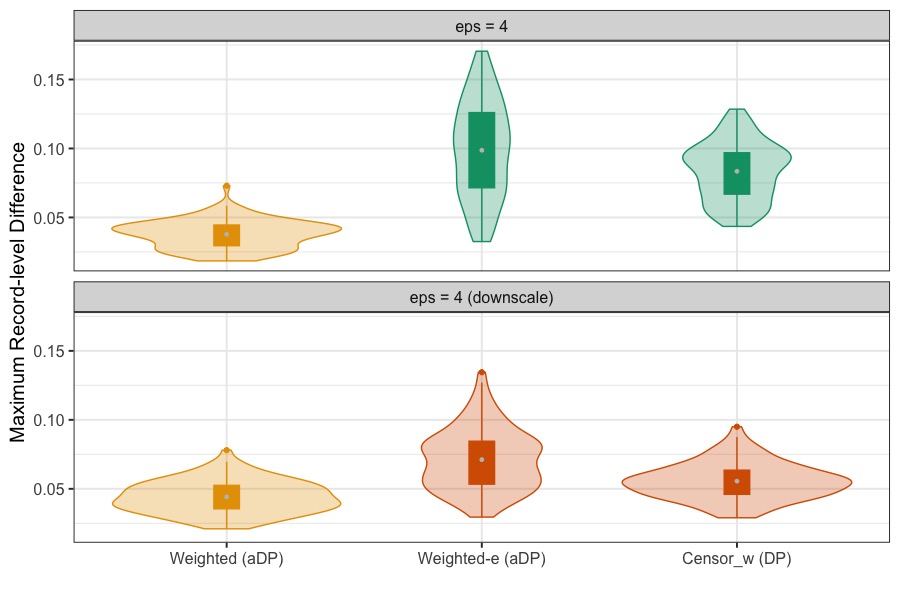}
     \caption{Violin plots of max-ECDF utility over $R = 100$ replicates, for the Weighted (aDP), the Weighted-e (aDP), and the Censor\_w (DP) at $\epsilon = 4$, without downscaling (top) and with downscaling (bottom).}
      \label{fig:sim:ECDF3}
\end{figure}

\begin{figure}[H]
  \centering
   \includegraphics[width=0.8\textwidth]{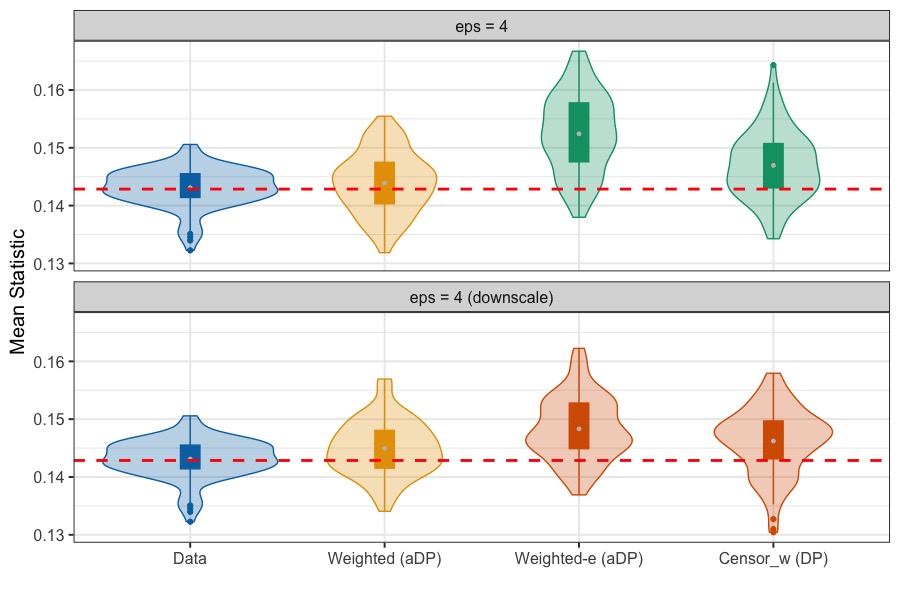}
      \caption{Violin plots of mean over $R = 100$ replicates for the Weighted (aDP), the Weighted-e (aDP), and the Censor\_w (DP) at $\epsilon = 4$, without (top) and with (bottom) downscaling. A dashed horizontal line at the analytical mean from Beta(0.5, 3) is included in each panel.}
      \label{fig:sim:mean_e4}
\end{figure}

\begin{figure}[H]
  \centering
   \includegraphics[width=0.8\textwidth]{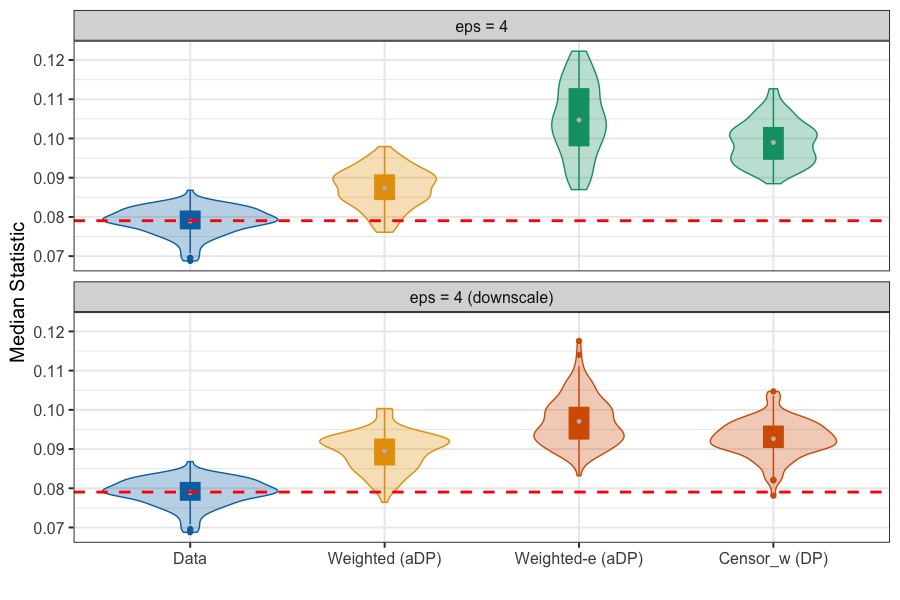}
      \caption{Violin plots of median over $R = 100$ replicates for the Weighted (aDP), the Weighted-e (aDP), and the Censor\_w (DP) at $\epsilon = 4$, without downscaling (top) and with downscaling (bottom). A dashed horizontal line at the analytical median from Beta(0.5, 3) is included in each panel.}
      \label{fig:sim:median_e4}
\end{figure}

\begin{figure}[H]
  \centering
   \includegraphics[width=0.8\textwidth]{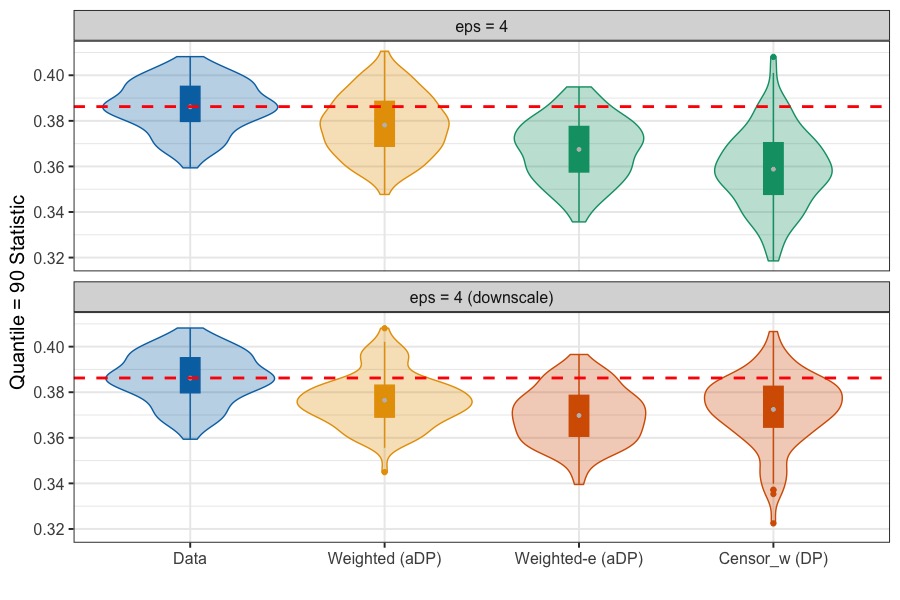}
     \caption{Violin plots of the 90th quantile over $R = 100$ replicates for the Weighted (aDP), the Weighted-e (aDP), and the Censor\_w (DP) at $\epsilon = 4$, without downscaling (top) and with downscaling (bottom). A dashed horizontal line at the analytical 90th quantile from Beta(0.5, 3) is included in each panel.}
      \label{fig:sim:q90_e4}
\end{figure}

\end{document}